\documentclass[11pt,a4paper,reqno]{amsart} 
\usepackage{amsfonts,amsthm, amscd, epsfig, amsmath, amssymb,enumerate}
\numberwithin{equation}{section}
\newtheorem{Th}{Theorem}[section]
\newtheorem{Le}[Th]{Lemma}

\newtheorem{Pro}[Th]{Proposition}

\newtheorem{definition}[Th]{Definition}
\theoremstyle{definition}

\newtheorem{remark}[Th]{Remark}

\DeclareMathSymbol{\leqslant}{\mathalpha}{AMSa}{"36} 
\DeclareMathSymbol{\geqslant}{\mathalpha}{AMSa}{"3E} 
\DeclareMathSymbol{\eset}{\mathalpha}{AMSb}{"3F}     
\renewcommand{\leq}{\;\leqslant\;}                   
\renewcommand{\geq}{\;\geqslant\;}                   
\newcommand{\dd}{\,\text{\rm d}}             
\newcommand{\suptwo}[2]{\sup_{\substack{#1 \\ #2}}} 
\newcommand{\inftwo}[2]{\inf_{\substack{#1 \\ #2}}} 
\newcommand{\sumtwo}[2]{\sum_{\substack{#1 \\ #2}}} 
\newcommand{\prodtwo}[2]{\prod_{\substack{#1 \\ #2}}}     

\makeatletter
\def\captionfont@{\footnotesize}
\def\captionheadfont@{\scshape}

\long\def\@makecaption#1#2{%
  \vspace{2mm}
  \setbox\@tempboxa\vbox{\color@setgroup
    \advance\hsize-6pc\noindent
    \captionfont@\captionheadfont@#1\@xp\@ifnotempty\@xp
        {\@cdr#2\@nil}{.\captionfont@\upshape\enspace#2}%
    \unskip\kern-6pc\par
    \global\setbox\@ne\lastbox\color@endgroup}%
  \ifhbox\@ne 
    \setbox\@ne\hbox{\unhbox\@ne\unskip\unskip\unpenalty\unkern}%
  \fi
  \ifdim\wd\@tempboxa=\z@ 
    \setbox\@ne\hbox to\columnwidth{\hss\kern-6pc\box\@ne\hss}%
  \else 
    \setbox\@ne\vbox{\unvbox\@tempboxa\parskip\z@skip
        \noindent\unhbox\@ne\advance\hsize-6pc\par}%
\fi
  \ifnum\@tempcnta<64 
    \addvspace\abovecaptionskip
    \moveright 3pc\box\@ne
  \else 
    \moveright 3pc\box\@ne
    \nobreak
    \vskip\belowcaptionskip
  \fi
\relax
}
\makeatother
\def\writefig#1 #2 #3 {\rlap{\kern #1 truecm
\raise #2 truecm \hbox{#3}}}

\newcommand{\cA}{\ensuremath{\mathcal A}}
\newcommand{\cB}{\ensuremath{\mathcal B}}
\newcommand{\cC}{\ensuremath{\mathcal C}}

\newcommand{\cE}{\ensuremath{\mathcal E}}
\newcommand{\cF}{\ensuremath{\mathcal F}}
\newcommand{\cG}{\ensuremath{\mathcal G}}
\newcommand{\cH}{\ensuremath{\mathcal H}}

\newcommand{\cL}{\ensuremath{\mathcal L}}

\newcommand{\cN}{\ensuremath{\mathcal N}}
\newcommand{\cO}{\ensuremath{\mathcal O}}

\newcommand{\cR}{\ensuremath{\mathcal R}}

\newcommand{\cU}{\ensuremath{\mathcal U}}
\newcommand{\cV}{\ensuremath{\mathcal V}}

\newcommand{\frH}{\ensuremath{\mathfrak H}}

\newcommand{\bbC}{{\ensuremath{\mathbb C}} }

\newcommand{\bbI}{{\ensuremath{\mathbb I}} }

\newcommand{\bbN}{{\ensuremath{\mathbb N}} }

\newcommand{\bbR}{{\ensuremath{\mathbb R}} }

\newcommand{\bbZ}{{\ensuremath{\mathbb Z}} }

\newcommand{\Om}{\Omega} 
\newcommand{\eps}{\epsilon} \newcommand{\var}{{\rm Var}}
\newcommand{\be}{\begin{equation}}
\newcommand{\bestar}{\begin{equation*}}
\newcommand{\la}{\label} \newcommand{\La}{\Lambda}
\newcommand{\grad}{\nabla} \newcommand{\si}{\sigma}
\newcommand{\Si}{\Sigma} \newcommand{\tb}{\widetilde b}
\newcommand{\al}{\alpha} \newcommand{\ket}[1]{\left\vert#1\right\rangle}
\newcommand{\bra}[1]{\left\langle #1\right\vert}
\newcommand{\braket}[2]{\left\langle #1 \vert #2 \right\rangle}

\newcommand{\scalar}[2]{\left\langle #1,#2 \right\rangle}
\newcommand{\scalarnun}[2]{{\left\langle #1,#2\right\rangle}_{\nu_n}}

\newcommand{\wt}{\widetilde} 
\newcommand{\wb}{\bar} 
\newcommand{\tz}{\tilde z}
\newcommand{\mul}{\mu^\lambda}
\newcommand{\tnu}{\tilde \nu}
\newcommand{\tLa}{\tilde\Lambda}
\newcommand{\tf}{\tilde f}
\newcommand{\tga}{\tilde \gamma}

\let\a=\alpha \let\b=\beta   \let\d=\delta  \let\e=\varepsilon
 \let\g=\gamma \let\h=\eta      \let\l=\lambda
            
  \let\s=\sigma    \let\th=\vartheta
  
\let\D=\Delta     \let\L=\Lambda 
     \let\Si=\Sigma 

\def\smallno{\smallskip\noindent}
\def\medno{\medskip\noindent}
\def\bigno{\bigskip\noindent}
\def\\{\hfill\break}
\def\acapo{\hfill\break\noindent}
\def\thsp{\thinspace}

\def\tthsp{\kern .083333 em}

\def\?{\mskip -10mu}
\def\indbox#1{\hbox to \parindent{\hfil\ #1\hfil} }

\newfam\msafam
\newfam\msbfam
\newfam\eufmfam
\def\hexnumber#1{%
  \ifcase#1 0\or 1\or 2\or 3\or 4\or 5\or 6\or 7\or 8\or
  9\or A\or B\or C\or D\or E\or F\fi}
\font\tenmsa=msam10
\font\sevenmsa=msam7
\font\fivemsa=msam5
\textfont\msafam=\tenmsa
\scriptfont\msafam=\sevenmsa
\scriptscriptfont\msafam=\fivemsa        
\edef\msafamhexnumber{\hexnumber\msafam}%
\mathchardef\restriction"1\msafamhexnumber16
\mathchardef\ssim"0218
\mathchardef\square"0\msafamhexnumber03
\mathchardef\eqd"3\msafamhexnumber2C
\def\QED{\ifhmode\unskip\nobreak\fi\quad
  \ifmmode\square\else$\square$\fi}            
\font\tenmsb=msbm10
\font\sevenmsb=msbm7
\font\fivemsb=msbm5
\textfont\msbfam=\tenmsb
\scriptfont\msbfam=\sevenmsb
\scriptscriptfont\msbfam=\fivemsb
\def\Bbb#1{\fam\msbfam\relax#1}    
\font\teneufm=eufm10
\font\seveneufm=eufm7
\font\fiveeufm=eufm5
\textfont\eufmfam=\teneufm
\scriptfont\eufmfam=\seveneufm
\scriptscriptfont\eufmfam=\fiveeufm

\def\({\left(}
\def\){\right)}
\let\Dir=\cE
\let\Z=\integer

\let\neper=e
\let\ii=i
\let\mmin=\wedge
\let\mmax=\vee

\let\id=\identity

\let\sset=\subset

\def\nep#1{ \neper^{#1}}

\def\tc{\thsp | \thsp}
\let\<=\langle
\let\>=\rangle

\def\Var{ \mathop{\rm Var}\nolimits }

\def\gap{\mathop{\rm gap}\nolimits}

\def\ninf#1{ \| #1 \|_\infty }

\outer\def\nproclaim#1 [#2]#3. #4\par{\medbreak \noindent
   \talato(#2){\bf #1 \Thm[#2]#3.\enspace }%
   {\sl #4\par }\ifdim \lastskip <\medskipamount 
   \removelastskip \penalty 55\medskip \fi}
\def\thmm[#1]{#1}
\def\teo[#1]{#1}
\def\sttilde#1{%
\dimen2=\fontdimen5\textfont0
\setbox0=\hbox{$\mathchar"7E$}
\setbox1=\hbox{$\scriptstyle #1$}
\dimen0=\wd0
\dimen1=\wd1
\advance\dimen1 by -\dimen0
\divide\dimen1 by 2
\vbox{\offinterlineskip%
   \moveright\dimen1 \box0 \kern - \dimen2\box1}
}
\begin{document}
\title[Energy gap in the XXZ model]
{Asymmetric diffusion and the energy gap above the 111 ground state of
the quantum XXZ model}
\date{June 25, 2001}
\author[P. Caputo]{Pietro Caputo} 
\address{Dip. Matematica, Universita' di Roma Tre, L.go S. Murialdo 1, 
00146 Roma, Italy}
\email{caputo\@@mat.uniroma3.it} 
\author[F. Martinelli]{Fabio Martinelli} 
\address{Dip. Matematica, Universita' di Roma Tre, L.go
S. Murialdo 1, 00146 Roma, Italy} 
\email{martin\@@mat.uniroma3.it}

\begin{abstract}
We consider the anisotropic three dimensional XXZ Heisenberg ferromagnet
in a cylinder with axis along the $111$ direction and boundary
conditions that induce ground states describing an interface orthogonal
to the cylinder axis. Let $L$ be the linear size of the basis of the
cylinder. Because of the breaking of the continuous symmetry around the
$\hat z$ axis, the Goldstone theorem implies that the spectral gap above
such ground states must tend to zero as $L\to \infty$. In \cite{BCNS} it
was proved that, by perturbing in a sub--cylinder with basis of linear
size $R\ll L$ the interface ground state, it is possible to construct
excited states whose energy gap shrinks as $R^{-2}$. Here we prove that,
uniformly in the height of the cylinder and in the location of the
interface, the energy gap above the interface ground state is bounded
from below by $\text{const.}L^{-2}$.  We prove the result by first
mapping the problem into an asymmetric simple exclusion process
on $\Z^3$ and
then by adapting to the latter the recursive analysis to estimate from
below the spectral gap of the associated Markov generator developed in
\cite{CancMart}. Along the way we improve some bounds on the equivalence
of ensembles already discussed in \cite{BCNS} and we establish an upper
bound on the density of states close to the bottom of the spectrum.

\vskip.8cm

\noindent
{\em 2000 MSC: 82B10, 82B20, 60K35}

\noindent
{\bf Key words and phrases}: XXZ model, quantum interface, 
asymmetric exclusion process, equivalence of ensembles, spectral gap. 
  

\end{abstract}

%

\maketitle

\newpage




\section{Introduction}

In recent years there has been a great deal of investigation of
the anisotropic spin $\frac{1}{2}$ XXZ Heisenberg model defined by
\be
\cH_{\L} = -\sumtwo{x,y\in \L:}{|x-y|=1} 
        \frac{1}{\D}\big(\, S_x^1 S_y^1 + S_x^2 S_y^2\,\big) +S_x^3
        S_y^3 \; + \text{boundary conditions}
\la{model}
\end{equation}
where $\L\sset \bbZ^d$ and $\D>1$ measures the anisotropy. Sometimes the
parameter $\D$ is expressed as $\D= (q+q^{-1})/2$, $0<q<1$, and the
classical Ising model is recovered in the limit $q\to 0$. We refer in
particular the reader to 
\cite{Alcarazetal,Bruno1,BrunoToma1,BrunoToma2,BrunoToma3,Matsui,BCNS} and 
\cite{NS}.

As it is well known, the XXZ model has two ferromagnetically
ordered translation invariant ground states, but also ground states that
describe domain walls between regions of opposite sign of the
spins. More precisely, for $d\ge 3$ and using a quantum version of the
Pirogov--Sinai theory \cite{BorFro}, it is possible to prove the
existence of low temperature states describing an interface orthogonal
to the {\tt 001} direction (a kind of Dobrushin state), provided that
$\D$ is large enough. Quite surprisingly, and this is one of the main
reasons for the increasing interest in such a model, the anisotropy is
able, under certain circumstances, to stabilize a domain wall against
quantum fluctuations even when, classically, thermal fluctuations are
too strong to allow for a stable interface.

This is indeed the case for the so called {\tt 11, 111, $\dots$}
diagonal interfaces. The Ising model is not expected to have a Gibbs
state describing a diagonal interface at low temperature because the zero
temperature configurations compatible with the natural geometry and
corresponding boundary conditions are enormously degenerate. A rigorous
proof of such a result is available so far only in the solid--on--solid
approximation thanks to results of \cite{Ken}. On the
other hand it has been shown independently in \cite{Alcarazetal} and
\cite{Gott} that an appropriate choice of the boundary conditions in
(\ref{model}) can lead to ground state selection that favour a diagonal
interface. Let us be a little bit more precise. For definiteness we set
$d=3$.  We then take as domain $\L$ a cylinder with basis of linear size
$L$, height $H$ and axis along the {\tt 111} direction. The state of the
system is described by vectors in the tensor product Hilbert space
$\frH=(\bbC^2)^{\otimes|\La|}$.  Fix $\D>1$ and define $
A(\Delta)=\frac12\sqrt{1-\Delta^{-2}}$. Boundary condition are then
introduced as follows. We let
\be
\cH_\La = \sum_{b\in\cB_\La}\cH_b\,,
\la{modelbis}
\end{equation}
where $\cB_\L$ is the set of oriented bonds of $\bbZ^3$ inside $\L$, the
single bond hamiltonians $\cH^q_b$ are given by
$$
\cH_b=-\Delta^{-1}\left(S^1_{x_b}S^1_{y_b} + S^2_{x_b}S^2_{y_b}\right)
-S^3_{x_b}S^3_{y_b} + A(\Delta)\left(S^3_{y_b} - S^3_{x_b}\right) + \frac14\,,
$$
and we write the bond $b$ as $b=(x_b,y_b)$ if $\ell_{y_b}> \ell_{x_b}$, 
where $\ell_x = x_1+x_2+x_3$, $x=(x_1,x_2,x_3)$ is a signed distance to the
origin.
Notice that the terms $A(\Delta)\left(S^3_{y_b} - S^3_{x_b}\right)$
cancel everywhere except at the two basis of the cylinder and that the
third component of the spin is a conserved quantity. The constant
$\frac{1}{4}$ is there in order to have $\cH_b \ge 0$.

The reason for the special choice of the coefficient $A(\D)$ comes
mainly from the one dimensional system (see \cite{BrunoToma1},
\cite{BrunoToma2} and \cite{Alcarazetal}). For $d=1$ ($L=1$ in our
language) and boundary coefficient $A(\D)$ the system enjoys a $SU_q(2)$
quantum group symmetry and the ground state degeneracy is equal to
$H+1$. If instead we take the boundary coefficient different from
$A(\D)$ then the degeneracy is lifted. Moreover, in complete analogy
with the exact computation of the ground state wave function of
(\ref{modelbis}) in $d=1$, one can show that, in each sector with
$\sum_x S^{3}_x = (2n-|\L|)/2$, 
$n=0,1,\dots,|\L|$, there exists a unique ground
state of $H_\L$, denoted by $\psi_n$, with zero energy
\cite{Alcarazetal}. More precisely, with the
convention that $\ket{1}$ and $\ket{0}$ stand for spin ``up'' and spin
``down'' respectively, the ground state $\psi_n$ can be written as
\be
\psi_n = \sumtwo{\al\in\Om_\La:}{N_\La(\al)=n}\psi_n(\a)\bigotimes_{x\in \L}\, 
         \ket{\a_x} 
\la{ground}
\end{equation}
where $\Om_\L:=\{0,1\}^{\L}$, $N_\L(\a) := \sum_{x\in \L}\a_x$ and 
\be
\psi_n(\a) = \prod_{x\in \L}\, q^{\ell_x\al_x}
\end{equation}
The square of the coefficients $\psi_n(\a)$ can be interpreted as the
statistical weights of a (non translation invariant) canonical Gibbs
measure for a lattice gas with $n$ particles described by the variables
$\{\a_x\}$. The typical configurations of such a measure form a sharply
localized (depending on $n$) interface orthogonal to the {\tt 111}
direction, separating a region almost filled with particles ($\a_x=1$)
from an almost empty region ($\a_x=0$). That justifies the name
``interface ground state'' for the vector $\psi_n$. Because of the
degeneracy of the ground states $\psi_n$, $n=0,1,\dots |\L|$, the
continuous symmetry given by rotation around the $z$--axis is broken and
therefore the spectrum above zero energy must be gapless in the
thermodynamic limit (see
\cite{Matsui}). That makes, in particular, any attempt to go beyond the
zero temperature case quite hard. To the best of our knowledge the only
model with a state describing a {\tt 111}-interface 
also at positive temperature 
is the Falicov--Kimball model  \cite{DMN}. 

\smallno
The structure of the low-lying
excitations above the interface ground states of (\ref{modelbis})
was recently studied in great detail in a series of
interesting papers \cite{BCNS,BCNS2,BCNS3}.
The main result in the above papers is that one can construct excitations
localized in a sub--cylinder of $\L$ of radius $R\ll L$ such that their
energy gap is smaller than $k\,R^{-2}$ for a certain constant
$k=k(q)$. Moreover, in an appropriate scaling, the energy spectrum of
such low--lying excitations coincides with the spectrum of the $d-1$
Laplacian on a suitable domain. An important ingredient 
in these works is an equivalence of ensembles result that can be roughly
described as follows. If we replace in (\ref{ground}) the weights
$\psi_n(\a)$ by their associated grand canonical weights obtained by
adding a suitably chosen constant chemical potential $\l:=\l(\L,n)$
and if we remove the condition $N(\a)=n$, we obtain a new vector that we
call grand canonical ground state and denote it by $\psi^\l$. Then, for
any local observable $X$ that commutes with the total third component of
the spin, the difference between the two averages
$\scalar{\psi_n}{X\psi_n}$ and $\scalar{\psi^\l}{X\psi^\l}$ vanishes
as $L\to \infty$.

\medno

Let us now discuss our results. As pointed out in \cite{BCNS} it is generally
believed that the energy of the lowest excitations in 
the {\tt 111}-cylinder $\L$ with height $H$ and 
basis of linear size $L$, is not only bounded from above 
but also from below by $O(L^{-2})$, uniformly in $H$. 
Our main contribution is a proof of this 
lower bound on the energy gap, see Theorem \ref{teorema}. 
We also give a proof of the corresponding upper bound
by making an ansatz similar to that of \cite{BCNS}. We should emphasize
that in contrast to \cite{BCNS} we do not have a detailed control of the
$q-$dependent prefactors in the estimates but rather focus on
the uniformity in $n$ (total third component of the spin) and $H$ (height 
of the cylinder). Another result of this paper concerns an estimate
on the density of states.
Namely, we consider vectors $\psi_n^f$ of the
form
$$
\psi_n^f =
\sumtwo{\al\in\Om_\La:}{N_\La(\al)=n}f(\a)\psi_n(\a)\bigotimes_{x\in
\L}\,\ket{\a_x}
$$ 
where $f$ is a local bounded function of the variables $\{\a_x\}_{x\in
\L}$ such that $\psi_n^f$ is orthogonal to $\psi_n$. Then, using the
lower bound on the spectral gap, we prove that the spectral measure
$\rho_f(E)$ associated to the vector $\psi^f_n$ satisfies $\rho_f(E) \le
k_\e E^{1-\e}$ for any $\e >0$ as $E\to 0$, uniformly in $n \neq
0,\,|\L|$ and in $\L$ (see
Theorem \ref{teorema1}). We believe that, in the above
generality, a linear behaviour near the bottom of the spectrum is the
correct one. Along the way we partially improve the equivalence of
ensembles results of \cite{BCNS} (see section 3) and we provide a
probabilistic proof of the known result
(\cite{BrunoToma2}) that the spectral gap for the linear chain XXZ is
uniformly positive (but our bound is very rough compared with that
of \cite{BrunoToma2}).

\medno

We now briefly describe our approach. 
Let $\frH_n$ denote the sector of
the Hilbert space $\frH$ with $\sum_{x\in \L}\a_x = n$ and 
define the normalized states
$$
\nu_n(\a) =
\frac{\psi_n^2(\a)}{\sum_\eta \psi_n^2(\eta)}\,.
$$
Using the positivity of the ground
states $\psi_n$ we may define a unitary transformation between $\frH_n$
and $L^2(\Om_\La,\nu_n)$ by formally multiplying 
by $\psi_n^{-1}$. 
This transforms $\cH_{\L,n}$, the restriction of $\cH_\L$ to $\frH_n$,
into a new operator $\cG_{\L,n}$ on
$L^2(\Om_\La,\nu_n)$. The latter turns out to be nothing but
the Markov generator of an asymmetric simple exclusion process in $\L$
that can be roughly described as follows. 
We have $n$ particles in $\La$ and each particle jumps to an
empty neighbouring site with rate proportional to $q$ if the signed
distance from the origin is increased (by one) and to $q^{-1}$ if it is
decreased.  The number of particles is a conserved quantity and by
construction the measure $\nu_n$ is reversible for the process since
$\cG_{\L,n}$ is self adjoint in $L^2(\Om_\La,\nu_n)$.  The spectral gap
of $\cG_{\La,n}$ coincides with the spectral gap of $\cH_{\L,n}$ and it
accounts for the smallest rate of exponential decay to equilibrium for
the above process in $L^2(\Om_\La,\nu_n)$. Note that the isotropic case
$q=1$ is the usual symmetric simple exclusion process.  Although we
discovered such an equivalence independently, we realized later on that
it was well known to physicists since some years
\cite{Alcaraz}. 

\smallskip

Once the problem has become a kind of reversible Kawasaki dynamics for a
classical lattice gas, we adapt to it some recent work
\cite{CancMart} (see also \cite{Lu-Yau} for a different approach) to
bound from below its spectral gap, recursively in $L$. Although our
asymmetric simple exclusion has certain advantages over a high
temperature truly interacting lattice gas because its grand canonical
measure is product, nevertheless several new problems arise,
particularly if one looks for results uniform in $n,H$, because of the
unboundedness of the signed distance $\ell_x$ entering in the canonical
measure $\nu_n$. 

\smallskip

As a final remark we observe
that all our results are restricted to spin $\frac12$. For higher spins one
can still compute exactly the ground state (see \cite{Alcarazetal}) for a
suitable choice of the boundary conditions and, as described above, it
is possible to unitarily transform the Hamiltonian into a Markov
generator. The interacting particle process one gets in this way is
however more involved than the one considered here. Particles of
different kind (namely different spin) appear and, besides the usual
asymmetric simple exclusion process, new transitions are allowed in
which pairs of particles of opposite spin are created or destroyed with
certain rates (see
\cite{Alcaraz}). We plan to analyze this new situation in a near future.

\medno

We conclude with a road map of the paper.
\begin{itemize}

\item In section 2 we fix the model, define the unitary transformation
leading to the Markov generator and state the main results.  

\item In section 3 we provide a series of technical tools including the
results on the equivalence of ensembles.
 
\item In section 4 we describe the recursive approach to prove the lower bound
on the spectral gap by assuming a key result that one may call
``transport theorem'' (see Theorem \ref{transport}). We also prove a
lower bound on the gap in one dimension uniformly in the number of
``up'' spins and in the height $H$.

\item In section 5 we prove the transport theorem. 

\item Finally in section 6 we prove the upper bound on the spectral gap
and the result on the spectral measure of local perturbations of the
ground state.

\end{itemize}

\subsection*{Acknowledgments} 
We warmly thank Bruno Nachtergaele and Pierluigi Contucci for 
enlightening discussions on their paper \cite{BCNS} and on the 
XXZ models in general.




\section{Setup and Main Results}

\subsection{Lattice, bonds, 111-planes, sticks and cylinders}
We consider the 3D integer lattice $\bbZ^3$, and
denote $e_i$, $i=1,\dots,3$ the unit vectors in the $i$-th direction. 
For any $x\in\bbZ^3$ we write $x_i=x\cdot e_i$ for the $i$-th coordinate of $x$
and denote by $\ell_x$ the signed distance from the origin
$$\ell_x=x_1+x_2+x_3\,.$$
A {\em bond} in $\bbZ^3$ is an oriented couple $b=(x,y)$, where $x,y\in\bbZ^3$ 
are neighbours, i.e.\ $\|x-y\|_1=1$ with $\|x\|_1=|x_1|+|x_2|+|x_3|$.
We denote ${\bbZ^3}^*$ the set of all bonds. 
A given $b\in{\bbZ^3}^*$ identifies two sites
$x_b,y_b\in\bbZ^3$ such that $b=(x_b,y_b)$. For
any subset $\La\subset\bbZ^d$ we call $\La^*$ the set 
of $b\in{\bbZ^3}^*$, such that $x_b,y_b\in\La$. 
For any $b$ we have $\ell_{x_b}-\ell_{y_b}=\pm 1$. 
We choose an orientation according to increasing values of $\ell$ and denote 
$$
\cB_\La=\big\{b\in\La^*:\,\;\ell_{y_b}=\ell_{x_b}+1\,\big\}\,.
$$
Given $h\in\bbZ_+$ we call $\cA_h$ the 111-{\em plane} at height
$h$, i.e.\
$$
\cA_h=\big\{x\in\bbZ^3:\,\;\ell_{x}=h\,\big\}\,.
$$
We define the infinite {\em stick} $\Si_\infty$ passing through the origin
as the doubly infinite sequence
$$
\dots,-e_1-e_2-e_3,-e_2-e_3,-e_3,0,
e_1,e_1+e_2,e_1+e_2+e_3,e_1+e_2+e_3+e_1,\dots
$$
We write $\Si_{x,\infty}$ 
for the infinite stick going through $x$, i.e.\ $\Si_{x,\infty}=x+\Si_\infty$.
Note that the union of $\Si_{x,\infty}$, 
$x\in\cA_0$ covers all of $\bbZ^3$.  
For every positive integer $H$ we define the finite stick 
$$
\Si_{H} = \big\{y\in\Si_\infty:
\quad \ell_{y}\in\{0,1,2,\dots,H-1\}\,\big\}\,.
$$
The finite stick through $x$ is then $\Si_{x,H}=x+\Si_H$.
When no confusion arises we shall simply write $\Si_x$ for a generic 
finite stick at $x$.
We will often consider cylindrical subsets of $\bbZ^3$ of the type
$$
\Si_{\Gamma,H}=\bigcup_{x\in\Gamma}\Si_{x,H}\,,\quad
\Gamma\subset\cA_0\,,$$ 
with some finite $\Gamma\subset\cA_0$, called the basis.
Then $\Si_{\Gamma,H}$ contains $H|\Gamma|$ sites, 
$|\Gamma|$ being the cardinality
of $\Gamma$. 
On the plane $\cA_0$ it is convenient to
parametrize sites as follows. Consider the 
two vectors $P_u=(1,-1,0)$ and $P_v=(0,1,-1)$. Then any $x\in\cA_0$ is
uniquely determined by a couple of integers $(x_u,x_v)$ with
$x=x_u P_u + x_v P_v$. We consider {\em tilted rectangles} 
$$R_{L,M}=\big\{x\in\cA_0:\;\;x_u\in\{0,1,\dots,L-1\}\,,
x_v\in\{0,1,\dots,M-1\}\,\big\}\,.$$
In this way $|R_{L,M}|=LM$. When $L=M$ we call $Q_L=R_{L,L}$ a {\em tilted
square}.
Corresponding cylinders $\Si_{Q_L,H}$ are denoted $\Si_{L,H}$.   
Note that there are no true neighbours on $\cA_0$. 
We say that two sites 
$x,y$ are neighbours in $\cA_0$ if $x,y\in\cA_0$ and
$|x_u-y_u|+|x_v-y_v|=1$. 

\subsection{Interface ground states of the XXZ model}
Consider a cylinder $\La=\Si_{\Gamma,H}$ for some $\Gamma\subset\cA_0$, $H\in\bbZ_+$.
The state of the system is described by vectors in the tensor product
Hilbert space $\frH=(\bbC^2)^{\otimes|\La|}$. 
Fix $q\in(0,1)$ and define $$\Delta=\frac12(q+q^{-1})\,,\;\;\;
A(\Delta)=\frac12\sqrt{1-\Delta^{-2}}\,.$$
The Hamiltonian operator is defined by
\be
\cH_\La = \sum_{b\in\cB_\La}\cH^q_b\,,
\la{ham}
\end{equation}
where single bond hamiltonians $\cH^q_b$, $b=(x_b,y_b)$ are given
\be
\cH^q_b=-\Delta^{-1}\left(S^1_{x_b}S^1_{y_b} + S^2_{x_b}S^2_{y_b}\right)
-S^3_{x_b}S^3_{y_b} + A(\Delta)\left(S^3_{y_b} - S^3_{x_b}\right) + \frac14\,.
\la{hambond}
\end{equation}
Here the spin operators $S_x^i$, $i=1,2,3$, are the Pauli matrices
\begin{equation*}
S^1_x=\begin{pmatrix}
0&1/2 \\ 1/2&0 \end{pmatrix}\,,\;\;\;
S^2_x=\begin{pmatrix}
0&-i/2 \\ i/2&0 \end{pmatrix}\,,\;\;\;
S^3_x=\begin{pmatrix}
1/2&0 \\ 0&-1/2 \end{pmatrix}\,.
\end{equation*}
Expressions (\ref{ham}) and (\ref{hambond}) give the usual
XXZ Hamiltonian, with the term proportional to $A(\Delta)$ 
accounting for boundary conditions which favour 111-interface states.
The term $1/4$ has been introduced so that ground states have zero
energy, see below.
 
We choose a basis for $\frH$ labeled by the two states
``up'' or ``down'' of the third component of the spin at each site, and
write it in terms of configurations 
$\al=\{\al_x\}_{x\in\La}$, with $\al_x\in\{0,1\}$ with the convention
that $\al_x=1$ stands for spin ``up'' while $\al_x=0$ stands for spin ``down''.
$\Om_\La=\{0,1\}^{\La}$ denotes the set of all configurations and
$\ket{\al}=\prod_{x\in\La}\ket{\al_x}$ stands for a generic basis vector.
For every $\varphi\in\frH$ we write
$$
\varphi(\al)=\braket{\al}{\varphi}\,.
$$
Since $\cH^q_b$ only acts on the bond $b$, a simple computation shows that
\be
\cH^q_b\ket{\al} = (q+q^{-1})^{-1}\left\{
q^{\al_{x_b}-\al_{y_b}}\ket{\al} - \vert\al^b\rangle\right\}\,,
\la{hambond2}
\end{equation}
where $\al^b := T_{x_b,y_b}\a$, and for a generic pair $x,y$,
$T_{x,y}\a$ denotes the configuration in which $\al_{x}$ and $\al_{y}$
have been exchanged,
$$(T_{x,y}\al)_{z} = 
\begin{cases}
\al_{y} & z=x\\
\al_{x} & z=y\\
\al_z & \text{ otherwise}
\end{cases}\,$$
In particular, formula (\ref{hambond2}) shows that if $\al=\al^b$,
then $\cH^q_b\ket{\al}=0$. Moreover, $\cH^q_b=\ket{\xi}\bra{\xi}$ is 
a projection onto the vector $\xi=\xi_b^q$ with 
$$
\xi(\al)=\frac1{\sqrt{1+q^2}}\left\{
q\al_{x_b}(1-\al_{y_b})-(1-\al_{x_b})\al_{y_b}\right\}\,.
$$
Let $\cN_\La$ denote the operator
\be
\cN_\La\ket{\al}= N_\La(\al)\ket{\al}\,,\quad N_\La(\al)=
\sum_{x\in\La}\al_x\,.
\la{nalfa}
\end{equation}
From (\ref{hambond2}) we see that $\cH_\La$ commutes with $\cN_\La$. 
We divide $\frH$ in $|\L|+1$ sectors corresponding to all possible values
of $N_\L$. Namely, given $\varphi\in\frH$ we write 
\begin{equation*}
\ket{\varphi}=\sum_{n=0}^{|\La|}\ket{\varphi_n}\,,\quad 
\ket{\varphi_n}=\sumtwo{\al\in\Om_\La:}{N_\La(\al)=n}\varphi(\al)
\ket{\al}\,.
\end{equation*}
In this way $\frH$ is unitarily equivalent to the direct sum
$\oplus_n\frH_n$, where $\frH_n$ is the closed subspace
of $\frH$ spanned by all vectors $\ket{\al}$ with $N_\La(\al)=n$. 
Now, ground states for the Hamiltonian (\ref{ham}) are vectors $\psi$ 
in $\frH$ such that $\cH_\La\ket{\psi} = 0$.
As in \cite{Alcarazetal}, \cite{BCNS} and \cite{BCNS2}, 
in each sector $\frH_n$, $n=0,1,\dots,|\L|$, there is a unique ground
state $\psi_n$ given by  
\be
\psi_n(\al)
=\begin{cases}\prod_{x\in\La}q^{\ell_x\al_x} & N_\La(\al)=n\\
0 & N_\La(\al)\neq n \end{cases}
\la{psin}
\end{equation}
We shall interpret $\psi_n^2$ as the weights of a {\em canonical} 
probability distribution
$\nu_n$ on $\Om_\La$, by writing 
\begin{equation*}
\nu_n(f) = \sum_{\al\in\Om_\La}\nu_n(\al)f(\al)\,,\quad f:\Om_\La\to\bbR\,,
\end{equation*}
with 
\be
\nu_n(\al)=\frac{\psi_n^2(\al)}{\sum_{\eta\in\Om_\La}\psi_n^2(\eta)}\,.
\la{nunal}
\end{equation}
It is convenient to introduce the corresponding {\em grand canonical} 
distributions. For every $\lambda\in\bbR$ we define the product measure
$\mu^\lambda$ on $\Om_\La$ given by
\begin{equation}
\mul(f) = \sum_{\al\in\Om_\La}\mul(\al)f(\al)\,,\quad \mul(\al)
=\prod_{x\in\La}
\frac{q^{2(\ell_x-\lambda)\al_x}}{1+q^{2(\ell_x-\lambda)}}\,.
\la{mula}
\end{equation}
For every $\lambda\in\bbR$, $\nu_n$ can be obtained from $\mul$
by conditioning on $N_\La(\al)=n$, i.e.\
\be
\nu_n = \mu^\lambda(\,\cdot\,|N_\La(\al)=n)\,.
\la{nunmula}
\end{equation}
To avoid confusion we sometimes write explicitly the 
region $\La$ we are considering and use  
the notations $\nu_{\La,n}$ and $\mu^\lambda_{\La}$
instead of $\nu_n$ and $\mu^\lambda$. We shall adopt the standard notation 
for the variance and covariances w.r.t.\ a measure $\mu$:
\be
\var_\mu(f)=\mu(f,f)=\mu\big((f-\mu(f))^2\big)\,,\quad
\mu(f,g)=\mu \big((f-\mu(f))(g-\mu(g))\big)\,.
\la{covas}
\end{equation}

\subsection{The spectral gap}
We call $\gap(\cH_\L)$ the energy
of the first excited state of $\cH_\La$. 
Let us write $\cH_{\La,n}$ for the restriction of $\cH_\La$ to 
the sector $\frH_n$. For each $n$ we define the gap
\be
\gap(\cH_{\La,n})=\inftwo{0\neq\varphi\in\frH_n:}
{\braket{\varphi}{\psi_n}=0}
\frac{\bra{\varphi}\cH_{\La,n}\ket{\varphi}}{\braket{\varphi}{\varphi}}\,.
\la{varprinn}
\end{equation}
We then have 
\be
\gap(\cH_\La)= \min_{n}\gap(\cH_{\La,n})\,.
\la{gammas}
\end{equation}

\subsection{Ground state transformation}
For each $n$ we consider now the Hilbert space 
$\widetilde\frH_n := L^2(\Om_\La,\nu_n)$ with scalar product
\begin{equation}
\scalarnun{\varphi}{\psi}=\sum_{\al\in\Om_\La}\nu_n(\al)
\overline{\varphi}(\al)\psi(\al)\,.
\la{scalarnun}
\end{equation}
The ground state transformation is defined by the unitary map
\bestar
\cU_n:\frH_n\to\widetilde\frH_n\,,\quad \varphi\to\cU_n\varphi\,,
\end{equation*}
where, for every $\al\in\Om_\La$ with $N_\La(\al)=n$,
\be
\big(\cU_n\varphi\big)(\al)=
(\nu_n(\al))^{-1/2}\varphi(\al)\,.
\la{unn}
\end{equation}
Let us define the operator $\cG_{\L,n}$ on $\widetilde\frH_n$ 
given by
\be
\cG_{\L,n} f(\al) = \frac{1}{(q+q^{-1})}\sum_{b\in\cB_\La}
\frac{\psi_n(\al^b)}{\psi_n(\al)}\left[f(\al^b)-f(\al)\right]\,.
\la{ln}
\end{equation}
A simple computation shows that 
$-\cG_{\L,n}$ is a symmetric, non-negative operator with
\be
\scalarnun{f}{(-\cG_{\L,n})f} = \frac{1}{2(q+q^{-1})}\sum_{\al\in\Om_\La}
\sum_{b\in\cB_\La}\nu_n(\al)\frac{\psi_n(\al^b)}{\psi_n(\al)}
\left|f(\al^b)-f(\al)\right|^2\,.
\la{flnf}
\end{equation}
Moreover, $\cG_{\L,n}${\bf 1}$=0$ with {\bf 1} 
denoting the constant $f\equiv1$.
We may define the gap in the spectrum of $-\cG_{\L,n}$ as
\be
\gap(\cG_{\L,n})=\inf_{0\neq f\perp {\bf 1}}
\frac{\scalarnun{f}{(-\cG_{\L,n})f}}{\scalarnun{f}{f}}\,.
\la{varprintilde}
\end{equation}  
Here the orthogonality $f\perp {\bf 1}$ means $\nu_n(f)=0$.
The next proposition motivates the introduction at this stage
of the operator $\cG_{\L,n}$ and of its spectral gap. 
\begin{Pro}
\la{gaps}
For every finite $\La\subset\bbZ^3$, for every $n=0,1,\dots,|\La|$, 
we have the identity
\be
\cH_{\La,n}=\cU_n^{-1}(-\cG_{\L,n})\cU_n\,.
\la{trans}
\end{equation}
In particular, $\gap(\cG_{\L,n})=\gap(\cH_{\L,n})$. 
\end{Pro}
\proof
If (\ref{trans}) holds we see that for any $\varphi\in\frH_n$,
\be
\bra{\varphi}\cH_{\La,n}\ket{\varphi} = 
\scalarnun{\cU_n\varphi}{(-\cG_{\L,n})\cU_n\varphi}\,.
\la{trans2}
\end{equation}
From this $\gap(\cG_{\L,n})=\gap(\cH_{\L,n})$ follows since
$$\braket{\varphi}{\psi_n}=0 \iff \nu_n(\cU_n\varphi)=0\,.$$
We turn to the proof of (\ref{trans}). 
Let $\widetilde\psi_n(\al) = \sqrt{\nu_n(\al)}$, so that 
$\varphi/\widetilde\psi_n=\cU_n\varphi$. 
Observe that for every $b\in\cB_\La$ we have $\ell_{y_b}=\ell_{x_b}+1$
and therefore
\be
\widetilde\psi_n(\al^b)=q^{\al_{x_b}-\al_{y_b}}\widetilde\psi_n(\al)\,.
\la{psialb}
\end{equation}
From (\ref{ham}) and (\ref{hambond2}) we see that
\begin{align}
\bra{\al}\cH_\La\ket{\varphi}&=
\frac{1}{(q+q^{-1})}\sum_{b\in\cB_\La}\left[q^{\al_{x_b}-\al_{y_b}}\varphi(\al)
- \varphi(\al^b)\right] \nonumber\\
&= \frac{1}{(q+q^{-1})}\sum_{b\in\cB_\La}\widetilde\psi_n(\al^b)
\left[\big(\cU_n\varphi\big)(\al) - \big(\cU_n\varphi\big)(\al^b)\right] 
\nonumber \\
&= \widetilde\psi_n(\al)\big[(-\cG_{\L,n})\cU_n\varphi\big](\al)
= \big[\cU_n^{-1}(-\cG_{\L,n})\cU_n]\varphi(\al)\,.
\la{hamfipsi}
\end{align}
Then (\ref{hamfipsi}) 
proves the claim. \qed

\subsection{Asymmetric exclusion process}
The operator $\cG_{\L,n}$ in (\ref{ln}) can be interpreted
as the generator of an interacting particle system
(see e.g.\ \cite{Li} for a general reference).
We define
\be
\grad_{xy}f(\al) := f(T_{x,y}\al)-f(\al)\,,\quad\quad
\grad_{b}f(\al):=\grad_{x_by_b}f\,.
\la{grad}
\end{equation}
Let also 
\be
c_{b}(\al)=\frac{q^{\al_{x_b}-\al_{y_b}}}{q+q^{-1}}\,,\quad b=(x_b,y_b)\,. 
\la{rates}
\end{equation} 
Then (\ref{ln}) may be rewritten
\be
\cG_{\L,n} f(\al) = \sum_{b\in\cB_\La}
c_{b}(\al)\grad_{b}f(\al)\,.
\la{ln2}
\end{equation} 
For every $n$, this defines a Markov Process with $n$
particles in $\La$ jumping to
empty neighbouring sites. The rate of a jump is proportional
to $q$ if a particle
moves from $\ell_x$ to $\ell_y=\ell_x+1$, 
and to $q^{-1}$ if it moves from $\ell_y$ to $\ell_x$. 
The number of particles is conserved and the measure
$\nu_n$ is reversible for the process since $\cG_{\L,n}$ is
self adjoint in $L^2(\Om_\La,\nu_n)$.

\smallskip
Consider a cylinder $\L:=\Si_{L,H}$ of height $H$
and whose 111-section $\Si_{L,H}\cap\cA_0$
is a tilted square $Q_L$ containing $L^2$ sites. Since
the degenerate cases $n=0$ and $n=HL^2$ are trivial
($\nu_n$ is simply a delta on the empty/full configuration), 
the variable $n$ will be 
assumed to range from $1$ and $HL^2-1$ in all statements below.
Our main results can be stated as follows.
\begin{Th}
\la{teorema}
For any $q\in(0,1)$
there exists a constant $k\in(0,\infty)$ such that 
for every positive integer $L$ of the form $L= 2^j$ for some
$j\in \bbN$ 
\begin{align}
\inf_{H,n}\gap(\cG_{\Si_{L,H},n})&\geq k^{-1} L^{-2}\,,
\la{gammabound} \\
\sup_{H,n}\gap(\cG_{\Si_{L,H},n})&\leq k L^{-2}\,.
\la{gammabound1}
\end{align}
\end{Th}
\begin{remark}
The proof of the lower bound 
(\ref{gammabound}) is based on a recursive analysis very
similar to that of \cite{CancMart}.  The problem here is in many aspects
simpler in view of the product structure of the underlying grand
canonical measure. On the other hand the approach of \cite{CancMart} has
to be modified in several important points because of the asymmetry in
the 111 direction.  In the next subsection we will reformulate the
problem in a more convenient fashion.  
\end{remark}

\noindent
The second result concerns the behaviour of the spectral measure
associated to suitable local functions near the bottom of the spectrum.
\begin{Th}
\la{teorema1}
In the same setting of Theorem \ref{teorema}, let $f$ be a bounded
function of zero mean w.r.t.\ $\nu_{n}$ and such that its support is
contained in a sub-cylinder $\L_0:=\Si_{L_0,H}$. Let $E_s$ denote the spectral
projection of the operator $-\cG_{\L,n}$ associated to the
interval $[0,s]$. Then, for any $q\in(0,1)$ and any $\eps >0$ there
exists a constant $k_\eps$ depending on $\eps$, $L_0$ 
and $\ninf{f}$ such that
\be
\sup_{\L,n}\scalar{f}{E_sf}_{\nu_{\L,n}} \le k_\eps \,s^{1-\eps}
\la{gammabound2} 
\end{equation}
\end{Th}
\noindent
\begin{remark}
Theorem \ref{teorema1} can be obviously formulated also for the
quantum Hamiltonian $\cH_{n}$ thanks to the unitary equivalence
stated in Proposition \ref{gaps}. In this context the result is as
follows. Consider the vector $\psi_n^f$ in the Hilbert space $\frH_n$
defined by $\psi_n^f(\a) = f(\a)\psi_n(\a)$, with $f$ as in the theorem. 
Then the spectral measure of $\psi_n^f$ has an almost linear bound close
to the bottom of the spectrum of $\cH_n$ 
\end{remark}

\subsection{From tilted to straight shapes} 
In order to avoid unnecessary complications coming from the
tilted geometry of our setting we shall make the following 
simple transformation which allows us to go from the 111-cylinders
described above to more familiar straight cylinders in $\bbZ^3$,
with axis along one coordinate axis. 
Recall that any point $x\in\bbZ^3$ is identified by the triple
$(x_u,x_v,\ell_x)$, where $\ell_x=x_1+x_2+x_3$ 
and the pair $(x_u,x_v)$ specifies the projection
of $x$ onto the $\cA_0$ plane, obtained as the intersection
$\Si_x\cap\cA_0$. We then have an isomorphism
$\Phi:\bbZ^3\to\bbZ^3$ given by
\be
(\Phi x)_1 = x_u\,,\quad (\Phi x)_2 = x_v\,,\quad (\Phi x)_3 = \ell_x
\la{mapT}
\end{equation}
The map $\Phi$ brings the 111-cylinder $\Si_{L,H}$ into the straight
cylinder $$\Phi\Si_{L,H}=\{0,1,\dots,L-1\}^2\times\{0,1,\dots,H-1\}\,.$$
We now introduce a new exclusion process, 
with $n$ particles in a given $\La\subset\bbZ^3$, 
jumping to empty neighbouring sites. 
A jump in the horizontal direction occurs with rate $1$ while
in the vertical direction it occurs with rate $q$ or $q^{-1}$ if
the particle is going upwards or downwards, respectively.  
The asymmetry of the original process along the 111 direction 
becomes here an asymmetry along the 001 direction (the third axis).
Consider the set of oriented bonds $\La^*$.
We choose an arbitrary orientation for the horizontal bonds, which we
denote $\cO_\La$.  $\cO_\La$ can be taken to be the set of
couples $b=(x,y)\in\La^*$ such that $x_3=y_3$, $y_1\geq x_1$
and $y_2\geq x_2$. For vertical bonds, which we denote $\cV_\La$,
we choose the orientation according to increasing values of
the third component. Thus 
$$
\cV_\La=\big\{b=(x,y)\in\La^*\,:\quad y_3= x_3+1\,\big\}\,.
$$ 
The generator of our new process can be written
\be
\cL_{\La,n} f(\al) = \sum_{b\in\cO_{\La}}\grad_{b}f(\al)
+ \sum_{b\in\cV_{\La}}q^{\al_{x_b}-\al_{y_b}}\grad_{b}f(\al)\,.
\la{gn}
\end{equation}
The generator $\cL_{\L,n}$ is symmetric in $L^2(\Om_{\La},\tnu_{\La,n})$,
where now $\tnu_{\La,n}$ is again given by (\ref{psin}) and (\ref{nunal})
but we interpret $\ell_x$ as the third coordinate $x_3$. 
The Dirichlet form associated to this process is defined by
\be
\cE_{\La,n}(f,f) = \cE_n^{\cO_\La}(f,f) + \cE_n^{\cV_\La}(f,f)\,, 
\la{eff}
\end{equation}
where $f:\Om_\La\to\bbR$ and 
\begin{align}
\cE_n^{\cO_\La}(f,f) = \frac12
\sum_{\al\in\Om_\La}\tnu_{\La,n}(\al)\sum_{b\in\cO_\La}
\big(\grad_bf(\al)\big)^2\,,\la{eoff} \\
\cE_n^{\cV_\La}(f,f) = \frac12 
\sum_{\al\in\Om_\La}\tnu_{\La,n}(\al)\sum_{b\in\cV_\La}
q^{\al_{x_b}-\al_{y_b}}\big(\grad_bf(\al)\big)^2\,.
\la{evff}
\end{align}
We have the following simple relation between the old process on $\La$
and the new process on $\Phi\La$.
Given $\La\subset\bbZ^3$ and $f:\Om_\La\to \bbR$,
set $\tLa=\Phi\La$ and $\tf(\al)=f(\al\circ\Phi^{-1})$.
\begin{Le}
\la{comparison}
For every $q\in(0,1]$ there exists $k<\infty$ such that
for every $\La$ and $n=0,1,\dots,|\La|$
\be
k^{-1}\scalarnun{f}{(-\cG_{\L,n})f}
\leq \cE_{\tLa,n}(\tf,\tf) \leq k \,\scalarnun{f}{(-\cG_{\L,n})f}
\la{straight}
\end{equation}
\end{Le}
\proof
To prove the second estimate observe that by (\ref{flnf})
\bestar
\cE_n^{\cV_{\tLa}}(\tf,\tf) \leq (q+q^{-1})
\scalarnun{f}{(-\cG_{\L,n})f}\,,
\end{equation*}
since for any $\tilde b \in\cV_{\tLa}$, $\tilde b=(x,y)$, 
we have $b:=(\Phi^{-1}x,\Phi^{-1}y)\in\cB_\La$ and the rates 
coincide apart from the factor $(q+q^{-1})$. 
Therefore we only have to control the horizontal part (\ref{eoff}).
Let us fix 
an horizontal bond 
$\tilde b\in\cO_{\tLa}$, $\tilde b=(x,y)$. The point is that 
$b:=(\Phi^{-1}x,\Phi^{-1}y)$ is not a true bond in $\cB_\La$
(since $\ell_{\Phi^{-1}x}=\ell_{\Phi^{-1}y}$),
but we can find $b_1,b_2\in\cB_\La$ such that,  
with the notation $T_b\al=\al^b$ one has 
$$
\al^{b} = T_{b_1}T_{b_2}T_{b_1}\al\,,\quad
\grad_{b}f(\al)=\grad_{b_1}f(T_{b_2}T_{b_1}\al) + 
\grad_{b_2}f(T_{b_1}\al) + \grad_{b_1}f(\al)\,.
$$
Observing that $c_b \geq q/(q+q^{-1})$ and that changes of measures 
give at most an additional factor $q^{-2}$, e.g.\
$\nu_n(\al)\leq q^{-2}\nu_n(T_{b_1}\al)$ for any $\al$, we can 
estimate
\begin{align*}
\sum_{\al\in\Om_{\tLa}} & \tnu_{\tLa,n}(\al)
\big(\grad_{\tilde b}\tf (\al) \big)^2 = 
\sum_{\al\in\Om_\La}\nu_{\La,n}(\al)
(\grad_{b}f(\al))^2 \\
& \leq 3 \sum_{\al\in\Om_\La}\nu_{\La,n}(\al)
\big[
(\grad_{b_1}f(T_{b_2}T_{b_1}\al))^2 + 
(\grad_{b_2}f(T_{b_1}\al))^2 + (\grad_{b_1}f(\al))^2 \big] \\
&\leq 3q^{-3}(q+q^{-1})\sum_{\al\in\Om_\La}
\nu_{\La,n}(\al)\big[ 
2c_{b_1}(\al) (\grad_{b_1}f(\al))^2 + 
c_{b_2}(\al)(\grad_{b_2}f(\al))^2
\big]\,.
\end{align*}
Summing over $\tilde b\in\cO_{\tLa}$ we obtain
$$\cE_n^{\cO_{\tLa}}(\tf,\tf)\leq 6q^{-3} (q+q^{-1})
\scalarnun{f}{(-\cG_{\L,n})f}\,.$$

\smallskip

\noindent
To prove the first inequality in (\ref{straight}) we repeat the 
same reasoning, observing that for every bond in 
$b\in \cB_\La$ either $b$ is along a
single stick in which case the bound is straightforward since
$\tilde b\in\cV_{\tLa}$ ($\tb$ is the image of $b$ under $\Phi$), 
or $b$ connects two different sticks. In the latter 
case there are $\tilde b_1\in\cO_{\tLa}$ and $\tilde b_2\in\cV_{\tLa}$ such that
the exchange across $b$ can be realized by successive exchanges
across $\tilde b_1$ and $\tilde b_2$ and the above arguments apply. 
\qed

\bigskip

\noindent
Thanks to Lemma \ref{comparison} we will obtain Theorem
\ref{teorema} as a consequence of the following
\begin{Th}
\la{thetheorem}
For any $q\in(0,1)$
there exists a constant $k\in(0,\infty)$ such that 
for every positive integer $L$ of the form $L=2^j$, $j\in
\bbN$
\begin{align}
\inf_{H,n}\gap(\cL_{\Si_{L,H},n})&\geq k^{-1} L^{-2}\,,
\la{bound}\\
\sup_{H,n}\gap(\cL_{\Si_{L,H},n})&\leq k L^{-2}\,.
\la{bound1}
\end{align}
\end{Th}

\smallskip

\begin{remark} Since there is complete symmetry between particles
($\al_x=1$) and holes ($\al_x=0$), for any $\La$
and any $n=0,1,\dots,|\La|$ we have
\be
\gap(\cL_{\La,n})=\gap(\cL_{\La,|\La|-n})\,.
\la{dual}
\end{equation}
\end{remark}

\bigskip

\noindent
{\bf Convention}. 
In the rest of the paper
we work in the straight geometrical setting described above.
With some abuse we keep all the notations unchanged and write
$\ell_x$ for the third coordinate $x_3$. In this way sets $\cA_h$
are now horizontal planes, $\Si_x$ denotes a vertical stick,
$\Si_{\Gamma,H}$ denotes a straight cylinder, $R_{L,M}$ denotes a rectangle
on the plane $\cA_0$ and so on. Moreover, 
the probability measure $\tnu_{n}$ will be simply written $\nu_n$, so that 
$\nu_n$ and $\mu^\lambda$ are defined as 
in (\ref{nunal}) and (\ref{mula}) provided $\ell_x$ stands for $x_3$.




\section{Preliminary Results}

\noindent
In this section we collect several preliminary technical results that
will enter at different stages in the 
proof of our two main results. As a rule, in what follows $k$ denotes a positive
finite constant depending only on $q$, whose value may change from line to line.

\subsection{Mean and variance of the number of particles}
In this first paragraph we give some elementary bounds on the statistics of the
number of particles in a stick and on the chemical potential as a
function of the mean number of particles. Part of the results discussed
below have already been derived with more accurate constants in
\cite{BCNS} in the case of an interface sitting roughly in the middle of
the cylinder. 
Here we need results that are uniform in the location of
the interface.  

\smallno

Let us consider a single stick of height $H$, $\L:=\Si_{H}$,
and the grand canonical measure $\mu^\l$ on $\L$, $\l\in\bbR$.
We have the following simple relations between $\l$ and 
$m(\l):=\mu^\l(N_{\L})$, the mean number 
of particles in $\L$.

\begin{Le}
\la{meannumber}
For each $q\in(0,1)$ there exists $k<\infty$ such that
for every $H\geq 1$
\begin{align}
 (\l - k)\vee \frac{\l}2
\leq m(\l) &\leq k+\l  &\text{ if } &\;\l\in[0,H-1]
\la{mn1}\\
\frac1k\,q^{2|\l|} \leq m(\l) &\leq k\,q^{2|\l|} &\text{ if } &\;\l < 0
\la{mn2}\\
H-k\,q^{2(\l-H)} \leq m(\l) &\leq H & \text{ if } &\;\l \geq H \la{mn3}
\end{align}
\end{Le}

\proof
From (\ref{mula}) we have the identity
\be
m(\l)=\sum_{j=0}^{H-1} \frac{q^{2(j-\l)}}{1+q^{2(j-\l)}}
\la{mn4}\,.
\end{equation}
If $\l>0$, the summand in (\ref{mn4}) is bounded below by $1/2$ for all
$j\leq [\l]$ and 
the bound $m(\l)\geq \l/2$ is straightforward. 
When $\l\in[0,H-1]$, writing $\l=[\l]+\{\l\}$ we have
\be
m(\l)=[\l]+1 -\sum_{k=0}^{[\l]}
\frac{q^{2(k+\{\l\})}}{1+q^{2(k+\{\l\})}} + \sum_{l=1}^{H-1-[\l]}
\frac{q^{2(l-\{\l\})}}{1+q^{2(l-\{\l\})}}\,.
\la{mn5}
\end{equation}
The estimate $|m(\l)-\l|\leq k$
then follows easily from (\ref{mn5}). This proves (\ref{mn1}). 
To prove (\ref{mn2}) observe that if $\l<0$
\be
m(\l)=q^{2|\l|}\sum_{j=0}^{H-1} \frac{q^{2j}}{1+q^{2(j-\l)}}\,,
\la{mn6}
\end{equation}
and therefore
$$q^{2|\l|}\frac{1-q^{2H}}{2(1-q^2)}
\leq m(\l) \leq q^{2|\l|}\frac{1-q^{2H}}{1-q^2}\,.$$
Finally (\ref{mn3}) follows from
\be
m(\l)=\sum_{j=0}^{H-1} \frac1{q^{2(\l-j)}+1}
= H - q^{2(\l-H)}\sum_{k=0}^{H-1}\frac{q^{2(k+1)}}{1+q^{2(k+1+\l-H)}}\,.
\la{mn7}
\end{equation}
\qed

\medno

Next we consider $\s^2(\l):=\var_{\mu^\l}(N_{\L})$, the variance of the 
number of particles in $\L$. 
\begin{Le}
\la{varnumber}
For each $q\in(0,1)$ there exists $k<\infty$ such that
for every $H\geq 1$, $\l\in\bbR$
\be
\frac1k\leq q^{2\left[\l\wedge 0 + (H-\l)\wedge 0\right]}\,
\s^2(\l)\leq k\,.
\la{vn}
\end{equation}
\end{Le}
\proof
Recall that 
\be
\s^2(\l)=\sum_{j=0}^{H-1}
\frac{q^{2(j-\l)}}{(1+q^{2(j-\l)})^2}=\sum_{j=0}^{H-1}
\left(q^{-(j-\l)}+q^{(j-\l)}\right)^{-2}\,.
\la{vn1}
\end{equation}
By symmetry
(particle-hole duality), it is sufficient to consider the
range $\l\leq H/2$. For the upper bound observe that (\ref{vn1}) 
is bounded above by
\be
\sum_{j=0}^{H-1} q^{-2|j-\l|}\leq
\begin{cases}
\frac{1+q^2}{1-q^2} & 0\leq\l\leq H/2
\\
q^{2|\l|}\frac1{1-q^2} & \l<0
\end{cases}
\la{vn2}
\end{equation}
For the lower bound we estimate (\ref{vn1}) below by
\be
\frac14 \sum_{j=0}^{H-1} q^{-2|j-\l|} \geq
\frac14\begin{cases}
q^2 & 0\leq\l\leq H/2\\
q^{2(|\l|+1)} & \l<0
\end{cases}
\la{vn3}
\end{equation}
\qed

\begin{remark}
\la{variance_vs_n}
A simple consequence of the bounds in Lemma \ref{meannumber} and Lemma
\ref{varnumber} is the following. Let $\L_L=\Si_{L,H}$ 
be the usual cylinder containing
$HL^2$ sites and let $\si^2_{L}(\l)$ denote the variance of $N_{\L_L}$ w.r.t.\
$\mu^\l$. Let also $m_L(\l)=\mu^\l(N_{\L_L})$. Since 
$m_L(\l)=L^2m(\l)$ and $\si^2_L(\l)=
L^2\si^2(\l)$ the estimates of the two lemmas 
can be combined to obtain that there exists $k=k(q)<\infty$ such that 
for any $\l\in\bbR$ and any $H\geq 1$
\be
\frac1k (m_L(\l)\wedge L^2) \leq \si^2_L(\l) \leq k\, (m_L(\l)\wedge L^2)\,.
\la{vnvsn}
\end{equation}
\end{remark}

\subsection{Comparison of canonical and grand canonical measures}
In this second paragraph we will discuss some simple but important
results on the equivalence and comparison between the finite volume
canonical measure $\nu_{\L,n}$ and the grand canonical one
$\mu_{\L}^\l$, where $\L :=\Si_{L,H}$ and the chemical potential
$\l=\l(\L,n)$ is chosen in such a way that $\mu_\L^\l (N_\L) =n$.
Although some of the results discussed below have already been discussed
in the seminal paper \cite{BCNS}, for later purposes we need to improve
the estimates obtained in \cite{BCNS} to get bounds similar to those
established in \cite{CM1} for general lattice gases. 
\acapo 
For notation convenience we drop all the
super/sub scripts in the measures.
 
\begin{Th}
\label{equiv}
For any $q\in (0,1)$ there exists a constant $k\in (0,\infty)$ such that 
for every positive integers $H,L$, $L_0\leq L$, for every $n=0,1,\dots,HL^2$ and for
any bounded function $f$ such that its support is contained in the
sub-cylinder $\Si_{L_0,H}$ we have
\begin{equation}
|\nu(f)-\mu(f) |\leq k\ninf{f}\frac{L_0^2}{L^2}
\la{equ1}
\end{equation}
\end{Th}

\begin{remark}
For the above result it is completely irrelevant whether we are in a tilted
or straight geometry. Also, because of horizontal translation invariance,
all what matters is that the support of $f$ is contained in some
cylinder with basis of linear size $L_0$. 
\end{remark}

\begin{remark}
It is interesting to compare our result with that of \cite{BCNS}. There
the dependence on $L_0,L$ is worst because the leading term is of the
order of $\frac{L_0^2}{L-L_0}$ but the coefficient in front of it,
$k\ninf{f}$ in our case, is better since it is proportional to $\sup_n
|\nu_{\L,n}(f)|$. On the other hand the proof
given below works in any dimension so that
the bound (\ref{equ1}) is valid for all $d\ge 2$ with $(L_0/L)^2$ replaced 
by $(L_0/L)^{d-1}$.
\end{remark}

\proof We begin by proving the result for $f(\a) = \a_x$, $x \in \Si_{L,H}$. 
In what follows $k$ will denote a generic constant depending only
on $q$ whose value may change from time to time. 
It will first be convenient to fix some additional handy notation. 
We let 
\begin{align*}
\s^2_y &:= \mu(\a_y,\a_y)\; 
       ;& \s^2&:= L^2 \sum_{y\in \Si_0} \s^2_y\; ;&  
\rho_x &:= \mu(\a_x)\;\\  \bar \a_x &:= \a_x-\rho_x\; ;& 
\b_{xy} &:= \frac{\s^2_y}{\s^2_x}\; ;&
 \phi_x(t) &:=
\mu(\nep{i\frac{t}{\s}\bar\a_x})\; 
\end{align*}
Notice that $\s^2 =\mu(N_\L, N_\L)$ because of the product
structure of the measure $\mu$.

Following \cite{CM1} we
begin by proving that for any $x,y \in\L$  
\be 
|\nu(\a_y - \b_{xy}\a_x-\mu(\a_y - \b_{xy}\a_x)| \leq k \frac{\s^2_y}{\s^2}
\la{equ2}
\end{equation}
for some constant $k = k(q)$. Once (\ref{equ2}) holds then, a summation
over $y$ together with the identity $\sum_y [\nu(\a_y)-\mu(\a_y)] =0$
and the definition of $\b_{xy}$ yields
\begin{align}
|\nu(\a_x)-\mu(\a_x)| \leq k\frac{\s^2_x}{\s^2 }\,,
\la{equ3}
\end{align}
which is a slightly stronger result than the sought bound
$\frac{k}{L^2}$ because 
\be
\s^2_x/\s^2 \leq \sum_{z\in
\Si_x}\s^2_z/\s^2 = \frac{1}{L^2}\,.
\la{equivx}
\end{equation}
In order to prove (\ref{equ2}) we write
\be
\nu(\a_y - \b_{xy}\a_x)-\mu(\a_y - \b_{xy}\a_x) = 
\frac{1}{2\pi \s \mu(N_\L=n)}\int_{-\pi\s}^{\pi \s}\dd t \, 
\mu(\nep{i\frac{t}{\s}(N_\L -n)}, \a_y - \b_{xy}\a_x)\,,
\la{equ3bis}
\end{equation}
where 
$$
\mu(N_\L=n)=\frac1{2\pi\s}
\int_{-\pi\s}^{\pi\s}\dd t \,\mu\bigl(\nep{i\frac{t}{\s}(N_\L-n)}\bigr)
$$
denotes the $\mu$-probability of having $n$ particles in $\L$.
Since $\mu$ is a product measure,
the absolute value of the numerator in the integrand is bounded from
above by
\be
\prod_{z\neq x,y} |\phi_z(t)|\, 
|\,\mu(\nep{i\frac{t}{\s}(\bar \a_x + \bar \a_y)}
[\bar \a_y - \b_{xy}\bar \a_x])|
\la{equ4}
\end{equation}
It is quite easy to check that
\be
|\phi_z(t)| \leq
\nep{-k\frac{\s_z^2}{\s^2} t^2}\qquad
\forall \; |t| \leq \pi \s
\la{equ5}
\end{equation}
for some constant $k = k(q)$, so that $\prod_{z\neq x,y} |\phi_z(t)| \leq
\nep{-k\, t^2}$. Moreover the identity $\nep{i\th} = 1 + i\th + R(\th)$
with $|R(\th)| \leq \frac{\th^2}{2}$ together with the definition of the
coefficient $\b_{xy}$ gives
\begin{align} 
|\,\mu(\nep{i\frac{t}{\s}(\bar \a_x + \bar \a_y)}
&[\bar \a_y - \b_{xy}\bar \a_x])|  \nonumber \\
&= |\,\phi_x(t)\mu(\nep{i\frac{t}{\s}\bar \a_y}
\bar \a_y) - \b_{xy}\phi_y(t)\mu(\nep{i\frac{t}{\s}\bar \a_x}
\bar \a_x)| \leq
k \frac{t^2}{\s^2} \s_y^2 
\la{equ6}
\end{align}
In conclusion, putting together (\ref{equ5}) and (\ref{equ6}), the
numerator of (\ref{equ3bis}) is bounded from above by
$k\frac{\s^2_y}{\s^2}$.

We are left with the analysis of the denominator in (\ref{equ3bis}).
We will show that 
\be
\s \mu(N_\L=n) \geq k^{-1} 
\la{equ7}
\end{equation}
uniformly in $n, L, H$. We have to distinguish between the case in which
$\s^2$ is ``large'' and the case in which $\s^2$ is ``small''. To be
more specific we fix a large number $A$ and start to analyze the case
$\s \ge A^{5}$. Again using Fourier analysis we write
\begin{align}
2\pi \s \mu(N_\L=n) &= 
\int_{-\pi\s}^{\pi\s}\dd t \,\mu\bigl(\nep{i\frac{t}{\s}(N_\L-n)}\bigr)
         \nonumber \\ &= \int_{A \leq |t| \leq \pi\s} \dd t
\,\mu\bigl(\nep{i\frac{t}{\s}(N_\L-n)}\bigr) +
\int_{-A}^{A}\dd t \prod_{x\in \L}\phi_x(t) \nonumber \\
&= \int_{A \leq |t| \leq \pi\s} \dd t
   \,\mu\bigl(\nep{i\frac{t}{\s}(N_\L-n)}\bigr)+ \int_{-A}^{A}\dd t
   \prod_{x\in \L} [1 - \frac{t^2}{2}\frac{\s_x^2}{\s^2} +
   R_x(t)]\nonumber \\ &= \int_{A \leq |t| \leq \pi\s} \dd t
   \,\mu\bigl(\nep{i\frac{t}{\s}(N_\L-n)}\bigr)+ \int_{-A}^{A}\dd t
   \prod_{x\in \L} [1 - \frac{t^2}{2}\frac{\s_x^2}{\s^2}] +
   \int_{-A}^{A}\dd t \, R(t) \nonumber \\
\la{equ8}
\end{align}
where $R_x(t) = \phi_x(t)- 1 + \frac{t^2}{2}\frac{\s_x^2}{\s^2}$ and 
a simple expansion gives
\be
R(t) = \sum_{j=1}^{|\L|}\frac{1}{j!}\sumtwo{x_1\dots x_j \in \L:}{x_1\neq x_2\neq\cdots
\neq x_j}
\prod_{i=1}^j R_{x_i}(t)\prod_{z\in \L\setminus \{x_1, \dots x_j\}} 
     [1 - \frac{t^2}{2}\frac{\s_z^2}{\s^2}] \,.
\la{equR}
\end{equation}
Let us examine the three
terms in the r.h.s.\ of (\ref{equ8}). The first one is smaller in absolute
value than
$k\nep{- k' A^2}$ because of the gaussian bound (\ref{equ5}). 
Observing that 
$$
\prod_{x\in \L} [1 - \frac{t^2}{2}\frac{\s_x^2}{\s^2}]\geq 
\nep{-\frac{t^2}{4\s^2}\sum_{x\in\L}\s_x^2} = \nep{-\frac{t^2}{4}}
$$
we see that
the second one is greater than $1/2$ provided that $A$ is
large enough, uniformly in all the parameters. 
Finally, using
$$
|R_x(t)| \leq k \frac{|t|^3}{\s^3}\s^2_x
$$ 
and (\ref{equR}), 
the absolute value of the third one is smaller than
\begin{align*}
2A \sup_{|t|\leq A}
\sum_{j=1}^{|\L|}\frac{1}{j!}&
\Big(\sum_{x\in \L} |R_x(t)|\Big)^j \,
\leq\,  
2A \sup_{|t|\leq A}
\sum_{j=1}^{|\L|}\frac{1}{j!}\Big(\frac{|t|^3}{\s}\Big)^j 
\, \leq\,  k\,A^{-1}\,.
\end{align*}
In conclusion, if $\s \ge A^{5}$ and $A$ is large enough (but
independent of $L,H,n$) (\ref{equ7}) holds true. 

\smallno
Let us now examine the
case $\s \leq A^{5}$ which, for large values of $L$ and $H$, corresponds
to an extremely low density of particles (cf.\ (\ref{vnvsn})) 
In this case we bound $\mu(N_\L=n)$ from
below as follows. If $L^2\leq 2n$ then we impose that all the particles in
the cylinder $\L$ are packed starting from the bottom and according to
an arbitrary ordering of the sites on each horizontal square $Q_L +
(0,0,\ell)$. It is not difficult to check that the probability of such
event is bounded below by $\exp{(-\gamma L^2)}$ for some $\gamma=\gamma(q)>0$,
uniformly in the height of the cylinder. But since $L^2\leq 2n\leq k\si^2$
(by (\ref{vnvsn})) we have a lower bound $\exp{(-k\gamma A^{10})}$. 
If instead $L^2 \ge 2n$ then we impose that all the particles are at
height $\ell=0$. The probability of this last event is equal to  
\be
\binom{L^2}{n}p_0^n (1-p_0)^{L^2-n} \prod_{\ell=1}^{H-1} (1-p_\ell)^{L^2}
\la{pois}
\end{equation}
where $p_\ell = \frac{q^{2(\ell-\l)}}{1+q^{2(\ell-\l)}}$ 
is the probability that
there is a particle at a site $x$ with $\ell_x=\ell$.  Notice that
$p_\ell \leq \frac{1}{2}$ for any $\ell\ge 0$ since $L^2 \sum_{\ell\ge 0} p_\ell =
n$. In particular
\be
\prod_{\ell=1}^{H-1} (1-p_\ell)^{L^2} \geq 
\exp\bigl(-4L^2\sum_{\ell=0}^{H-1} (1-p_\ell)p_\ell\bigr) 
= \exp(-4\s^2) \geq \exp(-4A^{10})\,.
\end{equation}
Finally, since $p_0 L^2 \leq n$ and $n\leq k\si^2\leq kA^{10}$, 
also the factor $\binom{L^2}n  p_0^n (1-p_0)^{L^2-n} $ is bounded away
from zero uniformly in $L, H$. 

\smallskip

We are now in a position to give the result in its full generality.
\acapo
For any $x\in \L_0=\Si_{L_0,H}$, set $\b_{x,f} := \mu(f,N_0)/\s^2_x$, where
$N_0$ denotes the number of particles in $\L_0$. Observing that
$$
\var_\mu(f)\leq 2\|f\|_\infty^2 \var_\mu(N_0)\,,
$$ 
and using Schwarz' inequality, it is not difficult to check that
\be
|\b_{x,f}| \leq k\ninf{f}\frac{\var_\mu(N_0)}{\s^2_x}\,.
\la{beta}
\end{equation} 
Thanks to (\ref{equ3}),
$$
|\b_{x,f}|\,|\nu(\a_x)-\mu(\a_x)| \leq k\ninf{f}\frac{\var_\mu(N_0)}{\si^2}
= k\ninf{f}\frac{L_0^2}{L^2}
$$
We can thus 
safely replace $f$ by $f-\b_{x,f}\a_x$.  \acapo
We proceed at this point exactly as in (\ref{equ3bis})$\cdots$(\ref{equ6}) 
and we observe that, because of the very definition of
$\b_{x,f}$ the analogous of (\ref{equ6}) holds, namely
\be
|\,\mu\big(\nep{i\frac{t}{\s}(\bar \a_x + \bar N_0)}
[f - \b_{x,f}\a_x]\big)| \leq k \ninf{f}\frac{L_0^2}{L^2}\,.
\end{equation}
The theorem then follows because of the bound (\ref{equ7}). \qed

\medno

Our second result provides a simple upper bound on the canonical
expectation of a rather general, non negative function $f$ in terms of
its grand canonical mean. The relevance of such bounds is that they
permit to control the canonical large deviations in terms of the grand
canonical ones (see \cite{CancMart} and especially \cite{CMR}).  
We begin with the very simple one dimensional case $\L= \Si_H$.
\begin{Pro}
\label{nu(x)}
For any $q\in (0,1)$ there exists a constant $k\in (0,\infty)$ such that
for any one dimensional stick $\L= \Si_H$, for every $n=0,\dots,H$
and for every $f\geq 0$ 
\begin{equation} 
\nu(f) \leq k \mu(f)
\end{equation} 
\end{Pro} 
\proof 
It suffices to observe that the
probability $\mu(N_\L=n)$ can be bounded from below by
\be
\mu(N_\L=n) \geq \prod_{x\in \L: \, \ell_x \leq n-1}\mu(\a_x)
                      \prod_{y\in \L: \, \ell_y \geq n}\mu(1-\a_y)
                  \geq k'(q) > 0\,,
\end{equation}
since, by Lemma \ref{meannumber}, there exists a constant $k=k(q)$
such that $|\l(n) -n| \leq k$. 
\qed

\smallno

Next we turn to the genuine three dimensional case $\L=\Si_{L,H}$.

\begin{Pro}
\label{canvar}
For any $q\in (0,1)$ and $\d \in (0,1)$ there exists a constant $k\in (0,\infty)$ such that 
for every positive integers $H$ and $L$, 
for every $n=0,1,\dots,HL^2$ and for every non--negative function $f$
whose support does not intersect more than $(1-\d)L^2$ sticks 
\begin{equation}
\nu(f) \leq k \mu(f)
\end{equation}
In particular, 
\begin{equation}
\nu(f,f) \leq k \mu(f,f)\,.
\end{equation}
\end{Pro}
\proof Let us denote by $\D_f$ the support of $f$. Then, using
(\ref{equ7}) together with the gaussian upper bound on the absolute
value of the characteristic function (\ref{equ5})
\begin{align} 
\nu(f) &= 
\frac{1}{2\pi \s \mu(N_\L=n)}\int_{-\pi\s}^{\pi \s}\dd t \, 
\mu(\nep{i\frac{t}{\s}(N_\L -n)}f) \nonumber \\
&\leq k \mu(f) \int_{-\pi\s}^{\pi \s}\dd t \, 
\prod_{x\in \L\setminus \D_f}|\phi_x(t)| \nonumber \\
&\leq k \mu(f) \int_{-\pi\s}^{\pi \s}\dd t \,\nep{-k\d t^2} \leq k' \mu(f) \,.
\la{ca1}
\end{align}
The result for the variance follows at once from
$$
\nu(f,f) \leq \nu\( (f-\mu(f)^2\) \leq k \mu(f,f)\,.
$$ 
\qed

\subsection{An estimate on covariances}
An important ingredient of our approach is the following version of a
well known estimate due to \cite{Lu-Yau} (see also \cite{CancMart}).
Set $\La=\Si_{L,H}$, and let $\nu$ denote the canonical
measure with $n$ particles in $\L$.
The result given below will be used in the recursive estimate of
section 4 in 
the regime of large $L$, see Theorem \ref{transport}. On the other hand
its proof uses an estimate for small values of $L$ 
(see (\ref{recursive2}) and (\ref{w(1)})) that will be proven 
independently later on. 

\smallskip

In the following $B$ denotes a 
planar section of $\La$, i.e.\  $B=\L\cap\cA_h$ for some integer $h\leq H-1$, 
and $N_B$ is the number of particles in $B$.
\begin{Pro}
\la{twoblock}
For every $q\in(0,1)$ and for every $\eps>0$ there exists $C_\eps = C_\eps(q) <\infty$ 
such that for any function $f$, any height
$H\geq 1$ and for all $n=0,\dots |\L|$ we have
\be
\nu(f,N_B)^2\leq 
\big(L^2\wedge n\big)\Big[\,C_\eps\,\Dir_\nu(f,f)+ \eps \Var_\nu(f)\,\Big]
\la{tb}
\end{equation}
\end{Pro}
\proof
We take $R\in\bbZ_+$ and write
the square $Q_L$ as the disjoint union of smaller squares $Q_R^i$ of
side $R\ll L$. 
This is no real loss since, in view
of the horizontal exchangeability of variables under $\nu$, the geometry 
of the basis does not play
any role and we can always assume $Q_L$ to be given by the union of
$\cup_iQ_R^i$ and $\bar Q$,  
where $\bar Q$ is a small region contained in a square of side $R$ which 
is inessential in the argument below. 
Let then $\L=\cup_i \Si_i$, 
$\Si_i:=\Si_{Q_R^i,H}$, $N_i:=N_{\Si_i}$ and let $\cF$ be
the $\si$-algebra generated by the random variables
$\big\{N_i\big\}$. For any pair of functions $f,g$ we have
\be
\nu(f,g)=\nu\big(\nu(f,g \tc \cF)\big)
+ \nu\big(f,\nu(g \tc \cF)\big)\,.
\la{tb1}
\end{equation}
Simple estimates then allow to write
\be
\nu(f,g)^2\leq 2\nu\big(\var_\nu(g \tc \cF)\big) 
                \nu\big(\var_\nu(f \tc \cF)\big)
             + 2 \var_\nu(f) \var_\nu\big(\nu(g\tc\cF)\big)\,.
\la{tb2}
\end{equation}
Now define the function $g=\sum_i g_i$, with 
$g_i=N_{B_i}-\beta N_i$, 
where $B_i=B\cap\Si_i$ and $\beta$ is a parameter to be fixed later on. 
Observe that with this choice
$\nu(f,N_B)=\nu(f,g)$ and 
\be
\var_\nu(g \tc \cF) = \sum_i\var_\nu(N_{B_i} \tc \cF_i)\,,
\la{tb3}
\end{equation}
where we used $\cF_i$ to denote the $\si$-algebra
generated by $N_i$. We fix now a value 
$n_i$ for $N_i$ and write $\l(n_i)$ for the corresponding 
chemical potential, i.e.\
\be
n_i = \mu^{\l(n_i)}(N_{i}) = |B_i|\sum_{\ell=0}^{H-1}
p_\ell^{\l(n_i)}\,,\quad 
p_\ell^{\l(n_i)} = \frac{q^{2(\ell-\l(n_i))}}{1+q^{2(\ell-\l(n_i))}} \,.
\la{tb4}
\end{equation}
Using Proposition \ref{canvar} we have (recall that $h$ is the level of 
every $B_i$)
\be
\var_\nu(N_{B_i}|N_{i}=n_i)\leq  k\var_{\mu^{\l(n_i)}}(N_{B_i})
= k|B_i|p_h^{\l(n_i)}(1-p_h^{\l(n_i)})\,,
\la{tb5}
\end{equation}
and by Lemma \ref{varnumber} and Lemma \ref{meannumber} we have
\be
\var_\nu(N_{B_i}|N_{i}=n_i)\leq  k \big(|B_i|\wedge n_i\big)\,.
\la{tb55}
\end{equation}
In particular, together with (\ref{tb3}) this implies 
\be
\max_{\{n_i\}:\;\sum_i n_i=n}\Var_\nu(g \tc \cF) \leq k\big(L^2\wedge n\big)\,.
\la{tb6}
\end{equation}
Since the measure $\nu(\cdot \tc \cF)$ is a product
$\otimes_i\nu(\cdot \tc \cF_i)$ and each factor $\nu(\cdot \tc \cF_i)$
satisfies a Poincar\'e inequality with a constant $W(R)$ uniform in
the conditioning field (see (\ref{recursive2}) and (\ref{w(1)})), we have
\begin{align}
\nu\left(\var_\nu(f \tc \cF)\right)
& \leq  \sum_i \nu\Big[\var_\nu(f \tc \cF_i)\Big] \nonumber\\
&\leq W(R) \sum_i \nu\Big[\sum_{b\in\Si_i^*}
\nu\big((\grad_bf)^2 \tc \cF_i\big)\Big] \nonumber\\
&\leq W(R)\, \Dir_\nu(f,f)
\la{tb7}
\end{align}
Plugging (\ref{tb6}) and (\ref{tb7}) in (\ref{tb2}) we obtain
\be
\nu(f,g)^2\leq 2kW(R)\,\big(L^2\wedge n\big)\Dir_\nu(f,f)
+ 2kL^2R^{-2}\, \var_\nu(f) \var_\mu\left(\nu(g_1 \tc \cF_1)\right)\,,
\la{tb8}
\end{equation}
where we have used again Proposition \ref{canvar} to bound 
$\var_\nu(\nu(g \tc \cF))$ in terms of 
$$
\var_\mu(\nu(g \tc \cF)) = L^2 R^{-2} \var_\mu\left(\nu(g_1 \tc \cF_1)\right)
$$ 
with 
$\mu :=\mu^\l$ the grand canonical measure on $\La$,
and $\l:=\l(n)$ such that $\mu(N_\L)=n$.
At this point we consider separately two cases corresponding to ``many'' 
and ``few'' particles respectively. 

\smallskip
We start with the case of  many particles.  
Suppose $n > \eps^2 L^2$. Here the claim (\ref{tb}) will follow from
\be
\var_\mu(\nu(g_1 \tc \cF_1))\leq k\,,
\la{tb9}
\end{equation}
by taking $R$ sufficiently large in (\ref{tb8}). 
To prove (\ref{tb9}) we begin by observing that by Theorem \ref{equiv}
$$
\sup_{n_1} \big|\nu(g_1 \tc N_1=n_1)-\mu_{\Si_1}^{\l(n_1)}(g_1)\big|
\leq k\,,
$$
and therefore it is sufficient to show 
\be
\var_\mu\left[\mu_{\Si_1}^{\l(N_1)}(g_1)\right] \leq k\,.
\la{tb12}
\end{equation}
An estimate on the variance of $\varphi(N_1):=\mu_{\Si_1}^{\l(N_1)}(g_1)$
can be obtained as follows. 
Since $\mu$ is a product $\otimes_{x\in\Si_i}\mu_x$, 
one has 
$$
\var_\mu(\varphi)\leq \sum_{x\in\Si_1}\mu\left(\var_{\mu_x}(\varphi)\right)\,.
$$
But 
\[
\mu\big(\var_{\mu_x}(\varphi)\big)=\si^2_x\,
\mu\Big(\big[\varphi\big(1+\sum_{y\neq x}\a_y\big)-
\varphi\big(\sum_{y\neq x}\a_y\big)\big]^2\Big)\,, 
\quad \si^2_x
= p_{\ell_x}^\l(1-p_{\ell_x}^\l)\,.
\]
It is not difficult now to deduce
\be
\var_\mu(\varphi)\leq k\Big[\sum_{x\in\Si_1}\si^2_x\Big]\, \mu\left( 
[\varphi(N_1 +1)-\varphi(N_1)]^2\right)\,.
\la{poincare}
\end{equation}
By Remark \ref{vnvsn} we know that 
$\sum_{x\in\Si_1}\si^2_x\leq k(R^2\wedge \mu(N_1))$.
Since $\mu(N_1)=nR^2/L^2$ we have
\be
\var_\mu\left(\varphi(N_1)\right)\leq 
k\,R^2 \mu\left[\big(\varphi(N_1+1)-\varphi(N_1)\big)^2\right]\,.
\la{tb13}
\end{equation}
Now we can choose $\beta$ so that
$$
\mu\left(\mu_{\Si_1}^{\l(N_1+1)}(N_{B_1})-\mu_{\Si_1}^{\l(N_1)}(N_{B_1})\right)=\beta\,.
$$
In this way the right 
hand side of (\ref{tb13}) is again a variance and a new application of 
(\ref{poincare}) gives
\begin{gather}
\var_\mu\left(\varphi\right)\leq
k\,R^4 \mu\left[(\Delta\varphi(N_1+1))^2\right]\,,\la{tb14}\\
\Delta\varphi(m):=\varphi(m+1)+\varphi(m-1)-2\varphi(m)\,.
\nonumber
\end{gather}
We are going to show that
\be
\sup_{m}|\Delta \varphi(m)|^2\leq k R^{-4}\,.
\la{tb15}
\end{equation}
We have
\be
\Delta \varphi(m)=
\int_0^1\int_0^1\left[\partial_t\partial_s\mu_{\Si_1}^{\l(m+s+t)}(N_{B_1})\right]
\dd t\dd s
\la{tb155}
\end{equation}
Set $\l_{s,t}= \l(m+s+t)$. Using the identities
\be
\partial_t\mu_{\Si_1}^{\l_{s,t}}(N_{B_1})=\partial_s\mu_{\Si_1}^{\l_{s,t}}(N_{B_1})=
\frac{\mu_{\Si_1}^{\l_{s,t}}(N_1,N_{B_1})}{\mu_{\Si_1}^{\l_{s,t}}(N_1,N_1)}\,,
\la{tb16}
\end{equation}
we have
\be
\partial_t\partial_s\mu_{\Si_1}^{\l_{s,t}}(N_{B_1})
= \frac{\mu_{\Si_1}^{\l_{s,t}}(N_1,N_1,N_{B_1})}
{\big(\mu_{\Si_1}^{\l_{s,t}}(N_1,N_1)\big)^2}
- \frac{\mu_{\Si_1}^{\l_{s,t}}(N_1,N_1,N_1)
\mu_{\Si_1}^{\l_{s,t}}(N_1,N_{B_1})}
{\big(\mu_{\Si_1}^{\l_{s,t}}(N_1,N_1)\big)^3}\,.
\la{tb17}
\end{equation}
Here we use the standard notation
$$
\mu(f,g,h) = \mu((f-\mu(f))(g-\mu(g))(h-\mu(h)))\,.  
$$
Direct computations show that for any $\l$ 
\begin{align*}
&\mu_{\Si_1}^{\l}(N_1,N_1) = |B_1|\sum_{j=0}^{H-1}p_j^\l(1-p_j^\l) \\
&\mu_{\Si_1}^{\l}(N_1,N_{B_1}) = |B_1|p_h^\l(1-p_h^\l)\\
&\mu_{\Si_1}^{\l}(N_1,N_1,N_1) = 
|B_1|\sum_{j=0}^{H-1}p_j^\l(1-p_j^\l)(1-2p_j^\l)\\
&\mu_{\Si_1}^{\l}(N_1,N_1,N_{B_1})=|B_1|p_h^\l(1-p_h^\l)(1-2p_h^\l)
\end{align*}
From (\ref{tb17}), using $|B_1|=R^2$ we have
\begin{gather}
\partial_t\partial_s\mu_{\Si_1}^{\l_{s,t}}(N_{B_1})
= R^{-2}\cC(h,H,\l_{s,t})
\la{tb18}\\
\cC(h,H,\l):=\frac{2p_h^\l(1-p_h^\l)\sum_{j=0}^{H-1}
p_j^\l(1-p_j^\l)[p_j^\l-p_h^\l]}
{\big(\sum_{j=0}^{H-1}p_j^\l(1-p_j^\l)\big)^3}\,.
\nonumber
\end{gather}
From (\ref{tb18}) and (\ref{tb155}) we see that (\ref{tb15}) will 
follow from
\be
\suptwo{h,H\in\bbN}{\l\in\bbR} \big|\cC(h,H,\l)\big| < \infty \,.
\la{tb19}
\end{equation}
A first estimate gives 
$$
\big|\cC(h,H,\l)\big|\leq 
\frac{2\sum_{j=0}^{H-1}
p_j^\l(1-p_j^\l)|p_h^\l-p_j^\l|}
{\big(\sum_{j=0}^{H-1}p_j^\l(1-p_j^\l)\big)^2}\,.
$$
Then Lemma \ref{varnumber} shows that 
$|\cC(h,H,\l)|$ is bounded whenever $\l\in[0,H-1]$. 
On the other hand 
if $\l\leq 0$ then $1-p_j^\l\geq 1/2$, whereas if 
$\l \geq H-1$ then $p_j^\l\geq 1/2$. 
In any case for $0\leq j,h \leq H-1$ 
$$
|p_h^\l-p_j^\l|\leq 2 \sum_{\ell=0}^{H-1}p_\ell^\l(1-p_\ell^\l)
$$
and (\ref{tb19}) follows.

\bigskip
We turn to analyze the case of few particles: $n \leq \eps^2 L^2$.  In
this case we simply take $R=1$ and call $\psi(N_1):=\nu(g_1 \tc
N_1)$. We may assume that $\eps^2k\leq \eps$.  
Thus looking back at (\ref{tb8}) we see that it will be
sufficient to show
\be
\var_\mu(\psi)\leq k\eps^2 \frac n{L^2}\,.
\la{tb90}
\end{equation}
Since now $\mu(N_1)=n/L^2$, (\ref{poincare}) gives
$$
\var_\mu(\psi)\leq k \frac n{L^2} 
\mu\left( [\psi(N_1 +1)-\psi(N_1)]^2\right)
$$
Choosing $\beta =\mu\left(\nu(N_{B_1}\tc N_1+1)-\nu(N_{B_1}\tc
N_1)\right)$, (\ref{tb14}) becomes
$$
\var_\mu(\psi)\leq k \frac {n^2}{L^4} 
\mu\big[(\Delta\psi(N_1+1))^2\big]\leq k\eps^2\frac n{L^2} 
\mu\big[(\Delta\psi(N_1+1))^2\big]\,.
$$
On the other hand a trivial bound (remember that now $B_1$ is just a
single site) gives $|\Delta\psi(m)|\leq 8$ and (\ref{tb90}) follows
immediately. \qed


\subsection{Glauber bound for the number of particles in half volume}
Consider the cylinder $\Si_{2L,H}$ and divide it into two parts
\be
\La=\Si_{2L,H}=\La_1\cup\La_2\,,\quad \La_1 = \Si_{R_{L,2L},H}\,,\;
\La_2=(L,0,0)+\Si_{R_{L,2L},H}\,.
\la{divide}
\end{equation}
Fix $n\in\{1,\dots,2L^2H=|\La|/2\}$, let $\nu:=\nu_{\La,n}$ denote the
canonical measure on $\La=\Si_{2L,H}$ with total particle number
$n$ and let $p_n(m)=\nu(N_{\La_1}=m)$.
We begin by establishing upper and lower bounds on the ratio
$\frac{p_n(m+1)}{p_n(m)}$ for $m\ge \frac{n}{2}$. In what follows
$\l_{s}$ will denote the chemical potential such that
$\mu^{\l_s}(N_{\L_1})=s$.
%
\begin{Le}
\la{sinm}
For any $q\in(0,1)$, there exists $k < \infty$ such that, 
uniformly in $L,H,n$ and $m\in [\frac{n}{2},n]$ 
\be
k^{-1}q^{\,2(\l_{m+1}-\l_{n-m-1})} \leq
\frac{p_n(m+1)}{p_n(m)} 
\leq k\, q^{\,2(\l_m-\l_{n-m})}\,.
\la{sinm1}
\end{equation}
Moreover, for every $\epsilon>0$ there exists $\delta\in(0,1)$
such that 
\be
\frac{p_n(m+1)}{p_n(m)} \leq \epsilon\,
\la{sinm10}
\end{equation}
whenever $m\in[\d n,n]$, uniformly in all other parameters.
\end{Le}
\proof
We write
\begin{align}
p_n(m+1) &= 
  \sumtwo{x\in \L_1}{y\in \L_2}\frac{\nu\big( \a_x(1-\a_y)\id_{N_{\L_1}=m+1}\big)}
{(m+1)(|\L_1|-n-m-1)}
      \nonumber \\
   &= 
     \sumtwo{x\in \L_1}{y\in \L_2} \frac{q^{\,2(\ell_x-\ell_y)} 
     \nu\big( \a_y(1-\a_x)\tc N_{\L_1}=m \big) p_n(m)}{(m+1)(|\L_1|-n-m-1)} 
      \nonumber \\
   &\leq k
       \sumtwo{x\in \L_1}{y\in \L_2} \frac{q^{\,2(\ell_x-\ell_y)} 
        \mu^{\l_m}\big((1-\a_x)\big) \mu^{\l_{n-m}}(\a_y)\,p_n(m)}{(m+1)(|\L_1|-n-m-1)} 
           \nonumber \\
&=
  k\Big[\frac{m}{m+1}\Big]\,\Big[\frac{|\L_1|-n-m}{|\L_1|-n-m-1}\Big]
        q^{\,2(\l_m-\l_{n-m})}\,p_n(m)
            \nonumber \\
&\leq k' q^{\,2(\l_m-\l_{n-m})}\,p_n(m)
\la{m+1/m}
\end{align}
where we used Proposition \ref{nu(x)} and the fact that $\nu(\cdot \tc
m)$ is the product of $\nu_{\L_1,m}$ and $\nu_{\L_2,n-m}$.
The lower bound in (\ref{sinm1}) 
can be obtained in a similar way if we write $p_n(m)$ as
$$
p_n(m) = 
  \sumtwo{x\in \L_1}{y\in \L_2}
\frac{\nu\big( \a_y(1-\a_x)\id_{N_{\L_1}=m}\big)}{(n-m)(|\L_1|-m)}
$$ 
and proceed as above.

\smallskip

In order to prove the estimate (\ref{sinm10}) one could use bounds
on the chemical potentials $\l_m, \l_{n-m}$. We prefer however a
different route and rewrite $p_n(m+1)$ as follows. For any $\ell$ we set
$A_\ell=\L_1\cap\cA_\ell$ and $B_\ell=\L_2\cap\cA_\ell$. Let also
$N_{A_\ell}(\a)$ be the number of particles in the plane $A_\ell$,
$V_{A_\ell}(\a)=2L^2-N_{A_\ell}(\a)$ the corresponding number of holes,
and similarly for $B_\ell$.  Then
\begin{align}
p_n(m+1) &=  
 \sum_{\ell=0}^{H-1}
   \sumtwo{x\in A_\ell}{y\in B_\ell}
   \nu\left(\frac{\a_x(1-\a_y)}{\sum_{h=0}^{H-1}
   N_{A_h}V_{B_h}}\id_{N_{\L_1}=m+1}\right)
              \nonumber\\
         & = \sum_{\ell=0}^{H-1}
   \sumtwo{x\in A_\ell}{y\in B_\ell}
   \nu\left(\frac{(1-\a_x)\a_y}{\sum_{h=0}^{H-1}
   (N_{A_h}+\d_{\ell,h})(V_{B_h}+\d_{\ell,h})} 
       \tc N_{\L_1}=m \right) p_n(m)
               \nonumber\\
         & \leq \nu\left(
    \frac{\sum_{\ell=0}^{H-1}V_{A_\ell}N_{B_\ell}}
     {\sum_{h=0}^{H-1}N_{A_h}V_{B_h}}
       \tc N_{\L_1}=m \right) p_n(m)\,.
               \la{sinm11}
\end{align}
On the event $N_{\L_1}=m,N_{\L_2}=n-m$ we have
$$
\frac{\sum_{\ell=0}^{H-1}V_{A_\ell}N_{B_\ell}}
     {\sum_{h=0}^{H-1}N_{A_h}V_{B_h}}
   = \frac{2L^2(n-m) - \sum_{\ell=0}^{H-1}N_{A_\ell}N_{B_\ell}}
 {2L^2m - \sum_{h=0}^{H-1}N_{A_h}N_{B_h}} \leq \frac{n-m}{2m-n}
   \leq \frac{1-\d}{2\d-1}
$$
for $\d n\leq m \leq n$. Therefore (\ref{sinm10}) follows
from the estimate (\ref{sinm11}).
\qed

\bigno
Next we establish a Poincar\'e inequality for the marginal of
$\nu$ on $N_{\La_1}$ with respect to the corresponding
Metropolis-Glauber dynamics. 
\begin{Pro}
\la{glauber}
For all $q\in(0,1)$, there exists a constant $k<\infty$ such that for
all integers $L,H$, for all $n\in\{0,1,\dots,4L^2H\}$ and for all
functions $g:\bbN\to\bbR$
\begin{equation}
\var_\nu
\big(g(N_{\La_1})\big)\leq
k\big(n\wedge L^2\big) \sum_m p_n(m)\wedge p_n(m+1)
\left[g(m+1)-g(m)\right]^2\,.
\la{glauber1}
\end{equation}
\end{Pro}
\proof
We follow closely the analogous result for translation invariant lattice
gases proved in \cite{CancMart} (see Theorem 4.4 there). 
Assume without loss of generality that $m \ge \frac{n}{2}$ and write
$p_n(m)$ as
$$
p_n(m) = \nep{-V_n(m)} \phi_n(m)
$$ 
where
\begin{align*}
     V_n(m) &:= 2\log(1/q)\,\big[\,m\,\l_m +(n-m)\l_{n-m}\,\big] - 
               \log\Big(\frac{Z^{m}}{Z}\Big)
            \\
     \phi_n(m) &:= \frac{\mu_{\L_1}^{\l_m}(N_{\L_1}=m) 
                  \mu_{\L_2}^{\l_{n-m}}(N_{\L_2}=n-m)}{\mu(N_\L=n)}
\end{align*}
Here the partition function $Z^{(m)}$ is given by
$$
Z^{(m)} := \prodtwo{x\in\L_1}{y\in \L_2}
                  \big(1+q^{2(\ell_x -\l_m)}\big)
                  \,\big(1+q^{2(\ell_y - \l_{n-m})}\big)
$$
and similarly for $Z$ but without the chemical potentials
$\l_m,\l_{n-m}$. \acapo
Were the factor $\phi_n(m)$ constant (better: of
bounded variation uniformly in $m,n$) then the sought
Poincar\'e inequality would follow at once, using e.g. Cheeger
inequality or Hardy's inequality \cite{MR}, from a convexity bound of
the form (see (\ref{derivative}) and (\ref{derivative2}) below)
$$
\frac{d^2}{dm^2} V_n(m) \ge k'(\d,q)\frac{1}{n\mmin L^2}
$$
Unfortunately the ratio $\frac{\phi(m)}{\phi(m')}$ can be rather large,
depending on $n$, if e.g.\ $m \approx n/2$ and $m'\approx n$. On the other 
hand Lemma \ref{sinm} shows that the
distribution $p_n(m)$ has at least exponential tails so that, as far as
the Poincar\'e inequality is concerned, the tails should be
irrelevant. That is indeed true and, according to section 4 of
\cite{CancMart}, the result follows if we can show that there exists
$\d <1$ with $1-\d \ll 1$ and a constant $k$ such that
\begin{align}
\sup_{m,m' \in [\frac{n}{2},\, \d n]}\; \frac{\phi(m)}{\phi(m')} &\le k 
\la{convesso1}\\
\sup_{\d n\, \le m\,\le n} \;\frac{p_n(m+1)}{p_n(m)} &\le \frac{1}{2} 
\la{convesso2}\\
\min_{m\in [\frac{n}{2},\, \d n]} \;\big[\,V(m+1)-V(m)\,\big] &\ge 
                                   \frac{1}{k(n\mmin L^2)}
\la{convesso3}
\end{align}
Inequality (\ref{convesso1}) follows at once from the fact,
proved in the discussion of the equivalence of ensembles, that
$\mu_{\L_1}^{\l_m}\big(N_{\L_1}=m\big)$ (and similarily for 
$\mu_{\L_2}^{\l_{n-m}}(N_{\L_2}=n-m)$) is comparable to the
inverse of the standard deviation of the number of particles in $\L_1$,
together with (\ref{vnvsn}).
\acapo
Inequality (\ref{convesso2}) is nothing but (\ref{sinm10}) above.
Finally (\ref{convesso3}) follows from the convexity of the
``potential'' $V_n(m)$. More precisely, since $V_n(m)$ is even
w.r.t.\ $\frac{n}{2}$, all what we need is
\be
\frac{d^2}{dm^2} V_n(m) = 2\log(1/q)\;\frac{d}{dm}\big[\,\l_m - \l_{n-m}\,]
\la{derivative}
\end{equation}
together with
\be
\frac{d}{dm}\l_m = \frac{1}{\log(1/q)} 
     \frac{1}{\Var_{\mu_{\L_1}^{\l_m}}(N_{\L_1})} 
\ge k'(\d,q)\frac{1}{n\mmin L^2} \qquad \forall m\in [\frac{n}{2},\, \d n]
\la{derivative2}
\end{equation}
and analogously for $\l_{n-m}$. Above we have used once more
(\ref{vnvsn}) to control the variance of the number of particles in
terms of its mean. \qed

\subsection{Moving particles}
In this paragraph we will show how to relate ``long jump terms'' of the form
$\nu_{\L,n}([\nabla_{xy}f]^2)$ with $x,y\in \L$ to sum of nearest
neighbor jumps along a path leading from $x$ to $y$. In what follows the
setting and the notation will be that of the preceding subsection, cf. \ (\ref{divide}).

\smallskip

We will analyze two different situations that we
call, for convenience, the many particles case {\tt (MP)} and the 
few particles case {\tt (FP)}. The definitions will depend on a 
parameter $\delta$
which will be forced to be
sufficiently small when needed (see the proof of Theorem \ref{transport}).  
\begin{itemize}
\item{
{\em Many particles}:
$$
H\geq 1\,,\;\d L^2\leq n\leq 2L^2H\,.\eqno({\tt MP})
$$
}
\item{
{\em Few particles}:
$$
H\geq 1\,,\;1\leq n\leq \d L^2\,.\eqno({\tt FP})
$$
}
\end{itemize}
In the {\tt MP} case let $A$ and $B$ be two horizontal
section of $\L_1$ and $\L_2$ at height $\ell_A$ and $\ell_B$
respectively, with $\ell_A \geq \ell_B$. 
In the {\tt FP} case, the sets $A$ and $B$ are instead given by
\be
A = \big\{x\in\La_1:\;\ell_x\leq h-1\big\}\,,\quad 
B = \big\{y\in\La_2:\;\ell_y=0\}\,.
\la{gk1.bis}
\end{equation}
where $h \in \bbN$ will be suitably tuned later on. 
Below we use $\nu(\cdot\tc m)$  for $\nu(\cdot\tc N_{\L_1}=m)$.
\begin{Pro}
\la{I1}
For any $q\in (0,1)$, any $n=1,\ldots \frac{|\L|}{2}$ and any $m\in
[\frac{n}{2},n]$
\begin{align}
\sumtwo{x\in A}{y\in B}q^{2(\ell_x-\ell_y)}
\nu & \big((\grad_{xy}f)^2\al_y(1-\al_x)\,\tc m\big)
\nonumber\\
& \leq C\Big\{ L^2 \sum_{b\in\cO_\La}
\nu\big((\grad_{b}f)^2\,\tc m\big)
+ \sum_{b\in\cV_\La}
\nu\big((\grad_{b}f)^2\,\tc m\big)\Big\}\,,
\la{2mp4}
\end{align}
where $C$ is a suitable constant depending on $q$ in the {\tt MP} case
and on $q,h$ in the {\tt FP} case.
\end{Pro}

The rest of this section is devoted to the proof of the above proposition.
For each couple of sites $x\in A,\,y\in B$, define a third site
$z=z(x,y)$ with $z_1=y_1,\,z_2=y_2$ and $\ell_z=\ell_x$. That is $z$ is
the unique element of $\cA_{\ell_x}\cap\Si_y$.  Since
$T_{xy}\al=T_{yz}T_{xz}T_{yz}\al$ we decompose $\grad_{xy}f$ in
\be
\grad_{xy}f(\al)=\grad_{yz}f(T_{xz}T_{yz}\al)
+ \grad_{xz}f(T_{yz}\al) + \grad_{yz}f(\al)\,.
\la{mp2}
\end{equation}
We then have two vertical moves corresponding to exchanges between $y$
and $z$, and one horizontal move corresponding to the exchange between
$x$ and $z$. Thus
\be
\sumtwo{x\in A}{y\in B}q^{2(\ell_x-\ell_y)}
\nu  \big((\grad_{xy}f)^2\al_y(1-\al_x)\,\tc m\big)
\leq  3
\left\{ \bbI_\cO + \bbI_\cV\right\}
\la{mp3}
\end{equation}
with 
\be
\bbI_\cO = 
\sumtwo{x\in A}{y\in B}q^{2(\ell_x-\ell_y)}
\nu\big((\grad_{xz}f(T_{yz}\al))^2
\al_y(1-\al_x) \tc m\big)
\la{mp4}
\end{equation}
and
\be
\bbI_\cV = 
\sumtwo{x\in A}{y\in B}q^{2(\ell_x-\ell_y)}
\nu\left[\big((\grad_{yz}f(T_{xz}T_{yz}\al))^2
+(\grad_{yz}f(\al))^2\big)\al_y(1-\al_x) \tc m\right]
\la{mp5}
\end{equation}
We analyze these terms separately.

\bigskip

\noindent
{\bf Vertical moves}. 
If we have a particle at $y$ and a hole at $x$ then 
$$ 
\big(T_{xz}T_{yz}\al\big)_y=\al_z\,,\quad
\big(T_{xz}T_{yz}\al\big)_x=\al_y=1\,,\quad
\big(T_{xz}T_{yz}\al\big)_z=\al_x=0\,.
$$
Computing $\grad_{yz}f(T_{xz}T_{yz}\al)$ we may thus assume $\al_z=1$
(it vanishes otherwise).  Since we have a particle both at $y$ and at
$z$ the change of variables $\al\to T_{xz}T_{yz}\al$ produces no extra
factors and
\begin{align}
\nu\big((\grad_{yz}f(T_{xz}T_{yz}\al))^2
\al_y(1-\al_x)\,\tc m\big)&= \nu\big((\grad_{yz}f(T_{xz}T_{yz}\al))^2
\al_y(1-\al_x)\al_z\,\tc m\big)\nonumber\\
&=\nu\big((\grad_{yz}f(\al))^2
\al_y\al_x(1-\al_z)\,\tc m\big)
\la{mp100}
\end{align}
For the second term in (\ref{mp5}) 
we have
$$
\nu\big((\grad_{yz}f(\al))^2
\al_y(1-\al_x)\,\tc m\big) = \nu\big((\grad_{yz}f(\al))^2
\al_y(1-\al_x)(1-\al_z)\,\tc m\big)\,,
$$
therefore (\ref{mp5}) becomes 
\begin{equation}
\bbI_\cV = 
\sumtwo{x\in A}{y\in B}q^{2(\ell_x-\ell_y)}
\nu\big((\grad_{yz}f(\al))^2\al_y(1-\al_z)\,\tc m\big)
\la{mp10}
\end{equation}
We need the following rather general result.  Given any $y\in\La$ and
$z\in\Si_y$ we write $\gamma_{zy}$ for the shortest path connecting $z$
and $y$ along the stick.
\begin{Pro}
\la{mp12}
For any $q\in(0,1)$, for any $y\in\La$ and $z\in\Si_y$ 
\be
q^{2[(\ell_z-\ell_y)\vee 0]}
\nu\big((\grad_{zy}f)^2\al_y(1-\al_z)\big)
\leq 4q^2(1-q^2)^{-1} 
\sum_{b\in\gamma_{zy}}\nu\big((\grad_bf)^2\big)
\la{mp120}
\end{equation}
\end{Pro}
\proof
Assume first that $\ell_z \geq \ell_y$. Let $M=\ell_z-\ell_y$ and
consider the sequence $y=x_0, x_1,\dots,x_M=z$, with
$\ell_{x_i}=\ell_{x_{i-1}}+1$.  Write $\al_i=\al_{x_i}$, $T_{i,i+1}$ for
the exchange operator $T_{x_ix_{i+1}}$ and $\grad_{i,i+1}$ for the
corresponding gradient.  We want to prove
\be
q^{2M}\nu\big((\grad_{0,M}f)^2\al_0(1-\al_M)\big)
\leq 4q^2(1-q^2)^{-1} 
\sum_{i=1}^{M}\nu\big((\grad_{i-1,i}f)^2\big)
\la{mp13}
\end{equation}
%
We have a particle at $x_0$ and a hole at $x_M$.
To compute $T_{0,M}$ we first bring the particle
from $x_0$ to $x_M$ and then bring the hole, which sits now 
at $x_{M-1}$, back to $x_0$. We write
\be
T_{0,M}\al = T_{0,1}T_{1,2}\cdots T_{M-1,M}T_{M-2,M-1}\cdots T_{1,2}T_{0,1}\al
\,.
\la{mp14}
\end{equation}
To fit the picture described above, formula (\ref{mp14}) should be read
backwards. The first part of the transformation is described by
operators
\be
R_0\al = \al\,,\quad R_i\al = T_{i-1,i}\cdots T_{1,2}T_{0,1}\al\,,\quad
i=1,2,\dots,M-1\,,
\la{mp15}
\end{equation}
while the second part is given by operators
\begin{align}
L_i\al & =T_{i,i+1}T_{i+1,i+2}\cdots 
T_{M-1,M}T_{M-2,M-1}\cdots T_{1,2}T_{0,1}\al\nonumber\\
&= T_{i,i+1}T_{i+1,i+2}\cdots 
T_{M-1,M} R_{M-1}\al\,,\quad i=1,2,\dots,M-1\,.
\la{mp16}
\end{align}
In this way a simple telescopic argument shows that 
\begin{align}
\grad_{0,M}f(\al) &= f(T_{0,M}\al)-f(\al) = 
f(L_{M-1}\al)-f(\al) + \sum_{i=1}^{M-1}\grad_{i-1,i}f(L_i\al)\nonumber\\
&=\sum_{i=1}^{M}\grad_{i-1,i}f(R_{i-1}\al)+
\sum_{i=1}^{M-1}\grad_{i-1,i}f(L_i\al)
\la{mp17}
\end{align}
Let us study these two contributions separately.
We start with the $R_i$'s. 
Observe that
\be
\big(R_i\al\big)_j=
\begin{cases}
\al_0     & j=i \\
\al_{j+1} & 0\leq j \leq i-1 \\
\al_j     & i+1\leq j\leq M
\end{cases}\quad\quad i=1,\dots, M-1\,.
\la{mp18}
\end{equation}
The change of variable $\al\to R_i\al$ produces then a factor
\be
r_i(\al)=\frac{\nu(\al)}{\nu(R_i\al)}
= q^{2\sum_{j=0}^{i-1}\ell_{x_j}\(\al_j-\al_{j+1}\)}
q^{2\ell_{x_i}\(\al_i-\al_0\)}\,.
\la{mp19}
\end{equation}
Writing $\ell_{x_j}=\ell_y+j$ we have
\begin{align}
\sum_{j=0}^{i-1}\ell_{x_j}&\(\al_j-\al_{j+1}\)
+\ell_{x_i}\(\al_i-\al_0\) \nonumber\\
&=  \sum_{j=0}^{i-1}j\(\al_j-\al_{j+1}\)
+i\(\al_i-\al_0\) = N_{[1,i]}(\al) - i \al_0\,,
\la{mp201}
\end{align}
where we have used the identity
\begin{align}
\sum_{j=0}^{i-1}j\(\al_j-\al_{j+1}\) &= 
\al_1-\al_2+2\al_2-2\al_3+\cdots + (i-1)\al_{i-1} - (i-1)\al_i
\nonumber \\
&= \(\al_1+\cdots + \al_{i-1}\) - (i-1)\al_i = N_{[1,i]}(\al)
- i\al_i\,,
\la{mp20}
\end{align}
and $N_{[1,i]}$ stands for the number of particles between
in $\{x_1,\dots,x_{i}\}$. 
Therefore we may estimate (\ref{mp19}) 
simply with $r_i(\al)\leq q^{-2i}$.
In particular,
\be
\nu
\left(\big(\grad_{i-1,i}f(R_{i-1}\al)\big)^2\right)
\leq q^{-2(i-1)} \nu
\left(\big(\grad_{i-1,i}f(\al)\big)^2\right)\,.
\la{mp202}
\end{equation}
On the other hand by Schwarz' inequality
\be
\left\{\sum_{i=1}^M\grad_{i-1,i}f(R_{i-1}\al)\right\}^2
\leq \sum_{j=1}^M q^{-2j}  
\sum_{i=1}^Mq^{2i}\big(\grad_{i-1,i}f(R_{i-1}\al)\big)^2\,.
\la{mp21}
\end{equation}
From (\ref{mp202}) and (\ref{mp21}) we arrive at
\begin{align}
q^{2M}\nu\left(\left\{
\sum_{i=1}^M\grad_{i-1,i}f(R_{i-1}\al)\right\}^2\right)&
\leq \sum_{j=1}^M q^{2(M-j)} 
\sum_{i=1}^M q^2 \nu
\left(\big(\grad_{i-1,i}f\big)^2\right)\nonumber\\
& \hskip-1cm \leq q^2(1-q^2)^{-1} \sum_{i=1}^M\nu
\left(\big(\grad_{i-1,i}f\big)^2\right)\,.
\la{mp22}
\end{align}

\smallskip

We turn to estimate the contribution of terms with $L_i$ in (\ref{mp17}).
Notice that
\be
\big(L_i\al\big)_j=
\begin{cases}
\al_M     & j=i \\
\al_0     & j=M \\
\al_{j+1} & 0\leq j \leq i-1 \\
\al_j     & i+1\leq j\leq M-1
\end{cases}\quad\quad i=1,\dots, M-1\,.
\la{mp23}
\end{equation}
The change of variable $\al\to L_i\al$ gives a factor
\be
l_i(\al)=\frac{\nu(\al)}{\nu(L_i\al)}
= q^{2\sum_{j=0}^{i-1}\ell_{x_j}\(\al_j-\al_{j+1}\)}
q^{2\ell_{x_i}\(\al_i-\al_M\)+2\ell_{x_M}\(\al_M-\al_0\)}\,.
\la{mp24}
\end{equation}
As in (\ref{mp20}) and (\ref{mp201}) we can write
\be
l_i(\al)=q^{2N_{[1,i]}(\al)}
q^{2(M-i)\al_M} q^{-2M\al_0}
\leq q^{-2M} q^{2N_{[1,i]}(\al)}\,.
\la{mp25}
\end{equation}
In particular,
\be
\nu
\left(q^{-2N_{[1,i]}(\al)}\big(\grad_{i-1,i}f(L_{i}\al)\big)^2\right)
\leq q^{-2M} \nu
\left(
\big(\grad_{i-1,i}f(\al)\big)^2\right)\,.
\la{mp251}
\end{equation}
Since we are assuming $\al_M=0$, we also have $(L_i\al)_i = \al_M=0$.
Thus in order to compute $\grad_{i-1,i}f(L_i\al)$ we may assume
$(L_i\al)_{i-1} = \al_i = 1$ and write directly
$\al_i\grad_{i-1,i}f(L_i\al)$.  Using again Schwarz' inequality
\be
\left\{\sum_{i=1}^{M-1}\al_i\grad_{i-1,i}f(L_{i}\al)\right\}^2
\leq \sum_{j=1}^{M-1} \al_j q^{2N_{[1,j]}(\al)}  
\sum_{i=1}^{M-1} q^{-2N_{[1,i]}(\al)}
\big(\grad_{i-1,i}f(L_{i}\al)\big)^2\,.
\la{mp26}
\end{equation} 
But
$$
\sum_{j=1}^{M-1} \al_j q^{2N_{[1,j]}(\al)}
= \sum_{j=1}^{N_{[1,M-1]}(\al)} q^{2j} \leq q^2(1-q^2)^{-1}\,.
$$
Now we can estimate as in (\ref{mp22}),
using (\ref{mp251}):
\begin{align}
q^{2M}\,\nu&\left(\left\{
\sum_{i=1}^{M-1}\grad_{i-1,i}f(L_{i}\al)\right\}^2
\al_0(1-\al_M)\right)\nonumber\\
&\leq q^2(1-q^2)^{-1}
\sum_{i=1}^{M-1}\nu
\left(\big(\grad_{i-1,i}f\big)^2\right)\,.
\la{mp27}
\end{align}
The estimates of (\ref{mp22}) and (\ref{mp27}) together 
with (\ref{mp17}) imply the claim (\ref{mp13}). 

\smallskip

It is not difficult to adapt the above argument to the case 
$\ell_z < \ell_y$. In this case, writing $M=\ell_y-\ell_z$, 
(\ref{mp13}) has to be replaced by 
\be
\nu\big((\grad_{0,M}f)^2\al_M(1-\al_0)\big)
\leq 4q^2(1-q^2)^{-1} 
\sum_{i=1}^{M}\nu\big((\grad_{i-1,i}f)^2\big)\,.
\la{mp139}
\end{equation}
Now $r_i(\al)\leq q^{2N_{[1,i]}(\al)}$ and $l_i(\al)\leq q^{2(M-i)}$,
so (\ref{mp139}) follows using the estimate (\ref{mp27}) for $R_i$-terms
and (\ref{mp22}) for $L_i$-terms. This ends the proof of
Proposition \ref{mp12}. \qed 

\bigskip

We can now go back to (\ref{mp10}) and continue the proof 
of Proposition \ref{I1}. Suppose first we are in case ({\tt MP}), 
i.e.\ the sets $A$ and $B$ are the planar sections at level
$\ell_A$ and $\ell_B$ respectively. Then, summing over
$x\in A$ in (\ref{mp10}),
\be
\bbI_\cV = 
2L^2q^{2(\ell_A-\ell_B)}\sum_{y\in B}
\nu\big((\grad_{yz}f(\al))^2\al_y(1-\al_z)\,\tc m\big)\,,
\la{2mp1}
\end{equation}
where $z$ is the unique element of $\Si_y\cap\cA_{\ell_A}$. 
Since $\ell_A\geq \ell_B$ using Proposition \ref{mp12} 
(with $\nu$ replaced by $\nu(\cdot\tc m)$) we easily obtain
\be
\bbI_\cV\leq 8q^2(1-q^2)^{-1}L^{2}
\sum_{b\in\cV_\La}
\nu\big((\grad_{b}f)^2\,\tc m\big) \,.
\la{mp11}
\end{equation}
Suppose now we are in the case ({\tt FP}). Here $A$ is a
sub-cylinder with height $h$ while $B$ is the planar section
at height $0$, cf.\ (\ref{gk1.bis}).    
Then from (\ref{mp10}), using Proposition \ref{mp12}
(with $\nu$ replaced by $\nu(\cdot\tc m)$), we see that 
\begin{align}
\bbI_\cV &= \sum_{x\in A}\sum_{y\in B} q^{2\ell_x}
\nu\big((\grad_{yz}f(\al))^2\al_y(1-\al_z)\,\tc m\big)
\nonumber\\
&\leq \sum_{x\in A} 4q^2(1-q^2)^{-1}\sum_{b\in\cV_\La}
\nu\big((\grad_{b}f)^2\,\tc m\big)\nonumber\\
&= 8q^2(1-q^2)^{-1}hL^2 \sum_{b\in\cV_\La}
\nu\big((\grad_{b}f)^2\,\tc m\big) \,.
\la{2mp2}
\end{align}

\bigskip

\noindent
{\bf Horizontal moves}. 
We go back to (\ref{mp4}). 
Observe that if there is a particle at $y$ and
a hole at $z$ the change of
variable $\al\to  T_{yz}\al$ produces the factor 
$q^{2(\ell_y-\ell_z)}$, thus canceling 
$q^{2(\ell_x-\ell_y)}$ in (\ref{mp4}).  We can estimate
\be
\bbI_\cO \leq 
\sumtwo{x\in A}{y\in B}
\nu\big((\grad_{xz}f)^2\,\tc m\big)
\la{mp6}
\end{equation}
Consider the case ({\tt MP}) first. Now both $x$ and $z$ lie
on the plane $\cA_{\ell_A}$. 
We fix a choice of paths on this plane as follows. 
For each couple $x,z\in\cA_{\ell_A}$ we take the path
$\gamma_{xz}$ obtained by connecting $x$ to $z$ first along
the direction $e_1$ and then along the direction $e_2$. 
As in the case of vertical moves we use a telescopic
sum to write $\grad_{xz}f$, thus obtaining two sums
over all bonds in the path $\gamma_{xz}$, cf.\ (\ref{mp17}). Since here we only have
horizontal exchanges there are no factors when we change variables
and we simply use Schwarz' inequality to obtain
\begin{align}
\nu\big((\grad_{xz}f)^2\,\tc m\big)&\leq 
2|\gamma_{xz}|
\sum_{b\in\gamma_{xz}}\nu
\big((\grad_bf)^2\,\tc m\big) \nonumber\\
&\leq 8L \sum_{b\in\gamma_{xz}}\nu
\big((\grad_bf)^2\,\tc m\big)\,,
\la{mp7}
\end{align}
where we used $|\gamma_{xz}|\leq 4L$.
Moreover, 
for any bond $b$ in the plane 
\be
\sumtwo{x\in A}{y\in B}1_{\{b\in\gamma_{xz}\}}\leq 4L^3\,.
\la{mp8}
\end{equation}
When we sum in (\ref{mp6}) 
we obtain
\be
\bbI_\cO \leq 32L^4
\sum_{b\in\cO_\La}
\nu\big((\grad_{b}f)^2\,\tc m\big)\,.
\la{mp9}
\end{equation}

Consider now the case ({\tt FP}). Here $A$ is the sub-cylinder
at height $h$ and $B$ is the planar section at height $0$, see
(\ref{gk1.bis}). The same estimate (\ref{mp9}) applies since
when summing over $x\in A$ we are now 
summing over all layers up to level $h-1$ and the r.h.s.\ in (\ref{mp9})
contains all bonds in such planes.

\bigskip

Collecting all the estimates in
(\ref{mp11}), (\ref{2mp2}) and (\ref{mp9}) and plugging
into (\ref{mp3}) we have obtained the desired bound (\ref{2mp4}). 
This completes the proof of Proposition \ref{I1}.




\section{Recursive Proof of Theorem \ref{thetheorem}}

We begin by describing the main ideas behind the recursive proof of
Theorem \ref{thetheorem}. 
Let $\var_{\La,n}(f)$ denote the variance of a function $f$ w.r.t.\ 
$\nu_{\La,n}$ and let
\be
W(\La) = \max_n \sup_{f}\frac{\Var_{\L,n}(f)}{\cE_{\La,n}(f,f)}\,,
\la{wla}
\end{equation}
where the supremum is taken over all non-constant 
$f: \Om_\La\to\bbR$.  When
$\La=\Si_{L,H}$ we write $W(L)$ for 
$\sup_{H}W(\Si_{L,H})$.  The lower bound
in Theorem \ref{thetheorem} follows if we can prove that for any
$q\in(0,1)$ there exists $k<\infty$ such that
\be
W(L)\leq kL^2 
\la{thebound}
\end{equation}
for any $L$ of the form $L= 2^j, \; j\in \bbN$.
In turn (\ref{thebound}) follows at once if we can prove that for any
$q\in(0,1)$ there exist $k<\infty$ and $L_0 >0$ such that
\begin{align}
W(2L) &\le 3 W(L) + k L^2 \qquad \; L\ge L_0 
                                    \la{recursive1}\\
W(2L) &\le k W(L) + k \qquad  \;\quad L\le L_0
                                    \la{recursive2}\\
& \qquad W(1) \leq k \,. \la{w(1)}
\end{align}

\subsection{Transport theorem and proof of the recursive inequalities}
The starting point to prove the recursive inequalities is the formula
for conditional variance that we now describe.
\acapo
Consider the cylinder $\Si_{2L,H}$ and divide it into two parts
(cf. (\ref{divide}))
$$
\La=\Si_{2L,H}=\La_1\cup\La_2\,,\quad \La_1 = \Si_{R_{L,2L},H}\,,\;
\La_2=(L,0,0)+\Si_{R_{L,2L},H}\,.
$$
Fix $n\in\{1,\dots,2L^2H=|\La|/2\}$ and let $\nu_{\La,n}$ denote
as usual 
the canonical measure on $\La=\Si_{2L,H}$ with total particle number $n$.
Conditioning on the number of particles in $\La_1$
decomposes the variance as follows:
\be
\var_{\La,n}(f)=\nu_{\La,n}\left(\var_{\La,n}\big(f\,|N_{\La_1}\big)\right)
+ \var_{\La,n}\left(\nu_{\La,n}\big(f\,|N_{\La_1}\big)\right)\,.
\la{decomp}
\end{equation}
Moreover, the above conditioning breaks $\nu_{\La,n}$ into the 
product $\nu_{\La_1,N_{\La_1}}\otimes\nu_{\La_2,n-N_{\La_1}}$, and therefore
\be
\nu_{\La,n}\left(\var_{\La,n}\big(f\,|N_{\La_1}\big)\right)
\leq \nu_{\La,n}\left(\var_{\La_1,N_{\La_1}}(f)+
\var_{\La_2,n-N_{\La_1}}(f)\right)\,.
\la{prodvar}
\end{equation}
The first term in (\ref{decomp}) is then estimated above using (\ref{wla}):
\begin{align}
\nu_{\La,n}\left(\var_{\La,n}\big(f\,|N_{\La_1}\big)\right)
&\leq W(\Si_{R_{L,2L},H})\,\nu_{\La,n}\left[\cE_{\La_1,N_{\La_1}}(f,f)
+ \cE_{\La_2,n-N_{\La_1}}(f,f)\right]\nonumber\\
& \leq W(\Si_{R_{L,2L},H})\,\cE_{\La,n}(f,f)
\la{I}
\end{align}
The analysis of the second term in (\ref{decomp}) is more delicate and
is directly related to transport of particles. In a sense it represents
the core of the proof. As we will see we will provide two different
bounds on the transport term: the first one is rather subtle but it is
valid only for large enough $L$. The second one, valid for any value of
$L$, is much more rough and therefore it will be used only for those
values of $L$ for which the first bound is not known to hold. For
simplicity, in what follows, we will always refer to these two
situations as the ``large'' or ``small''
$L$ case.  \acapo
Recall now the definition (\ref{eoff}) and (\ref{evff}) of the
horizontal and vertical part of the Dirichlet form. Then we have
\vfill
\eject
\begin{Th}
\la{transport}
\noindent
\begin{enumerate}[{\tt (i)}]
\item{\tt Large $L$.}\acapo
For any $\eps>0$, $q\in(0,1)$, there exists a finite constants
$C_\eps=C(\eps,q)$, $k=k(q)$ and $L_0=L_0(\eps,q)$ such that for any
$L>L_0$, $H\geq 1$ and for any $n=1,2,\dots,|\La|-1$ 
\begin{gather}
\var_{\La,n}\left(\nu_{\La,n}\big(f\,|N_{\La_1}\big)\right)
\leq \nonumber \\
k\left\{L^2\cE_n^{\cO_\L}(f,f) + \cE_n^{\cV_\L}(f,f)\right\}
+ C_\eps\,\cE_n(f,f) + \eps \,\var_{\La,n}(f)\,, \nonumber
\end{gather}
\item{\tt Small $L$. } \acapo
For any $q\in(0,1)$ and for any $L\ge 1$ there exists a finite constant
$C=C(L,q)$ such that, for any $H\geq 1$ and any $n=1,2,\dots,|\La|-1$
\begin{equation*}
\var_{\La,n}\left(\nu_{\La,n}\big(f\,|N_{\La_1}\big)\right)
\leq C\Big\{\,\cE_n(f,f) + 
\nu_{\La,n}\big(\Var_{\La,n}\big(f \tc N_{\La_1}\big)\big)\,\Big\}
\end{equation*}
\end{enumerate}
\end{Th}
\bigno
Once Theorem \ref{transport} is proven, we use (\ref{decomp}) and
(\ref{I}) to obtain
\begin{align}
\var_{\La,n}(f)&\leq 
(1-\eps)^{-1}
\left\{W(\Si_{R_{L,2L},H}) + k L^2\right\}\cE_{\La,n}(f,f)
\qquad \text{\tt for $L\ge L_0(\eps,q)$} \la{recursive2a}\\
\var_{\La,n}(f)&\leq 
\left\{ C_0\, W(\Si_{R_{L,2L},H}) + C_0\,\right\}\cE_{\La,n}(f,f)
\hskip 1,6cm \text{\tt for $L\le L_0(\eps,q)$} \la{recursive2b}
\end{align}
where $C_0=C(L_0,q)$.
In the large $L$ case (\ref{recursive2a}) proves in particular that
\be
W(\Si_{R_{2L,2L},H})\leq (1-\eps)^{-1}\big\{W(\Si_{R_{L,2L},H}) + k L^2
\big\}\,.
\la{dec2}
\end{equation}
We repeat now the decomposition (\ref{decomp}) for $\La=\Si_{R_{L,2L},H}$,
writing the latter cylinder as $\La=\La_1\cup\La_2$, 
$\La_1=\Si_{L,H}$ and
$\La_2=(0,L,0)+\Si_{L,H}$. Applying the same reasoning as above
we arrive at 
\be
W(\Si_{R_{L,2L},H})\leq (1-\eps)^{-1}\left\{W(\Si_{L,H}) + k L^2\right\}\,, \qquad
\forall \; H\geq 1 \;\text{ and } L\geq L_0(\eps,q) 
\la{dec3}
\end{equation}
From (\ref{dec2}) and (\ref{dec3}) we finally obtain
\be
W(2L)\leq (1-\eps)^{-2}\big\{W(L) + k L^2\big\}
 \qquad
\forall \; L\ge L_0(\eps,q) 
\la{dec4}
\end{equation}
which proves (\ref{recursive1}) due to the arbitrariness of $\eps$.
Equation (\ref{recursive2}) is proved similarly starting from
(\ref{recursive2b}). The bound (\ref{w(1)}) is given in the next subsection.

\subsection{Spectral gap in the one dimensional case}
In this final paragraph we prove that $W(1) <\infty$. In other words we
show that the spectral gap for the one dimensional asymmetric simple
exclusion process with generator (\ref{gn}) in the interval
$\Si_H:=\{x=(0,0,l):\, 0\le l \le H-1\}$ 
is bounded away from zero uniformly in
the number of particles and in the length of the interval $H$.  
Such a result has already been proved for the one dimensional XXZ model in
\cite{NS} but we decided to present a ``probabilistic'' proof for
completeness.  Here is our formal statement. 
Below we write $\nu:=\nu_{\,\Si_H,n}$, $\cE:=\cE_{\Si_H,n}$

\begin{Th}
For any $q\in (0,1)$ 
there exists a constant $k$ such that for any $H \geq 1$, any $n \leq
H$ and any function $f$
$$
         \Var_{\nu}(f) \le k\, \cE(f,f)\,.
$$
In particular, $W(1)\leq k$.
\end{Th}
\proof
Let $\g(n,H)$ denote the inverse spectral gap for the process in
$\Si_H$ with $n$ particles and let $\g(n) = \sup_H
\g(n,H)$. Notice that, by the particle--hole duality, $\g(n,H) =
\g(H-n,H)$ and therefore we will always assume, without loss of
generality, that $n \leq \frac{H}{2}$.  
If $n=1$ then it is well known, by e.g. Hardy's \cite{MR} or Cheeger
inequality
\cite{SC}, that $\g(1) <\infty$. Our idea is to perform a
sort of induction on the number of particles. For this purpose, for each
configuration $\a$ with $n$ particles we denote by $\xi:=\xi(\a)$ the
position of the last particle, namely $\xi = \max\{x\in\Si_H: 
\, \a_x=1\}$, and
we set $\rho(x)=
\nu(\xi =x)$ the probability that $\xi=x$. 
It is not difficult to see that the distribution of $\xi$ has an
exponential falloff so that, in particular, it satisfies a Poincare\'e
inequality with constant depending only on $q$.  More precisely we have
the following
\begin{Le}
For any $q\in (0,1)$ there exists $k$ such that for any $f(\a) :=
F(\xi(\a))$ 
\be
\Var_{\nu}(f) \le k \sum_{x\ge n-1} 
\big(\rho(x)\mmin \rho(x+1)\big) [F(x+1)-F(x)]^2 \qquad \forall \, H\geq n\,.
\end{equation}
\la{le1d}
\end{Le} 
\proof 
Using Cheeger inequality it is enough to 
prove that there exists $x_0 \ge n-1$ and $\b <1$ depending only on $q$
such that $\frac{\rho(x+1)}{\rho(x)}
\le \b$ for any $x\ge x_0$.
A simple change of variables (see (\ref{rho}) below) shows that 
\be
\frac{\rho(x+1)}{q\rho(x)}\nu\bigl((1-\a_x)\tc \xi =x+1 \bigr) =1
\end{equation}
In order to complete the proof it is enough to prove that 
$\nu\bigl(\a_x\tc \xi =x+1 \bigr)$ tends to zero for large $x$
uniformly in $n \le \frac{H}{2}$.
For any $x \ge n$ we have
\begin{align}
\nu\Bigl(\a_x\tc \xi =x+1\Bigr) 
       \,&\le\,\frac{\mu(\a_x=1\,;\,\a_{x+1}=1\,;\, \a_y=0 \;\forall
       y>x+1)}{\mu(N_{[0,n-2]}=n-1\,;\,\a_{x+1}=1\,;\,\a_y=0 \;\forall
       y\geq n-1, y\neq x+1)}
\nonumber \\
&\le \frac{\mu(\a_x)}{\bigl[\prod_{x=0}^{n-2}
      \mu(\a_y)\bigr]\,\bigl[\prodtwo{y\neq x:}{y\geq n-1}
\mu(1-\a_y)\bigr]} 
      \,\le k\, q^{2(x-n)}
\la{6d}
\end{align}
for some constant $k=k(q)$. Above we have used the explicit product
structure of the measure $\mu := \mu^{\l(n)}$ together with the fact proved
in (\ref{meannumber}) that $|\l(n)-n| \le k'$. 
\qed

\bigskip

We are now in a position to prove the theorem. 
We write
\be
\Var_{\nu}(f) = \nu\bigl(\Var_{\nu}(f \tc \xi )\bigr) +
\Var_{\nu}\bigl(\nu(f\tc \xi)\bigr) 
\la{1d}
\end{equation}
The first term in the r.h.s.\ of (\ref{1d}) coincides with 
$$
\sum_{x\geq n-1}\rho(x)\Var_{\nu_{[0,x-1],n-1}
\otimes\nu_{[x,H-1],0}}(f)
$$
and therefore it can be bounded from above, using the
definition of $\g(n,x)$, by
\be
\sum_{x\ge n-1} \rho(x)[\g(n-1,x)\mmin \g(x-n+1,x)] 
\Dir_{[0,x-1],n-1}(f,f)
\la{2d}
\end{equation}
because of the holes--particles duality. Here and below $\g(0,x)=0$
for all $x$.  
\acapo 
Let us examine the second term. 
Here we apply Lemma \ref{le1d} to
write
\be
\Var_\nu\bigl(\nu(f\tc \xi)\bigr) \le 
k \sum_{x\geq n-1} \rho(x)\mmin \rho(x+1)
\bigl[F(x+1)-F(x)\bigr]^2
\la{3d}
\end{equation}
where $F(x)=\nu(f\tc \xi=x)$. In order to compute the ``gradient''
of $F(x)$ we write
\begin{align}
F(x) &= \sum_{\a; \, \xi(\a)=x} 
\frac{\nu(\a)}{\rho(x)} f(\a)\a_x \nonumber \\
     &=\frac{\rho(x+1)}{q\rho(x)}
         \sum_{\a; \, \xi(\a)=x+1} 
         \frac{\nu(\a)}{\rho(x+1)} f(\a^{x,x+1})(1-\a_x)  
                 \nonumber \\
     &=\frac{\rho(x+1)}{q\rho(x)}\Big[
         \nu\Bigl(
           [\grad_{x,x+1}f](1-\a_x)\tc \xi =x+1\Bigr) + 
        \nu\Bigl(f,(1-\a_x)\tc \xi =x+1\Bigr)\Big] + 
       \nonumber \\
     &\quad + \frac{\rho(x+1)}{q\rho(x)}
                  \nu(f\tc \xi=x+1)
               \nu\big((1-\a_x)\tc \xi =x+1 \big)
\la{4d}
\end{align}
Setting $f=1$ gives  
\be 
\frac{\rho(x+1)}{q\rho(x)}\nu\bigl((1-\a_x)\tc \xi =x+1 \bigr) =1
\la{rho}
\end{equation}
Therefore the last term in the r.h.s.\ of (\ref{4d}) is equal to
$F(x+1)$. \acapo Thanks to (\ref{6d})
$\(\frac{\rho(x+1)}{q\rho(x)}\)
\le k$ uniformly in $n,H$. In conclusion
\begin{gather}
\bigl[F(x+1)-F(x)\bigr]^2 \le \nonumber \\  
k'\nu\Bigl(
           (\grad_{x,x+1}f)^2 \tc \xi =x+1\Bigr) 
+ k'\e(x)\nu\Bigl(f,f\tc \xi =x+1\Bigr)
\la{5d}
\end{gather}
where 
\be
\e(x) := \nu\Bigl((1-\a_x),(1-\a_x)\tc \xi =x+1\Bigr) \le k\, q^{2(x-n)}
\la{6dbis}
\end{equation}   
because of (\ref{6d}).
Thus the r.h.s.\ of (\ref{1d}) is bounded from above by
\be
\sup_{n-1\leq x\leq H-1}\Bigl[ 
\big\{\big[(\g(n-1,x)\mmin \g(x-n+1,x)\big](1 + 
       k\,q^{2(x-n)})\big\}\mmax k'\, \Bigr] 
\Dir(f,f)
\la{7d}
\end{equation} 
In other words we have proved the recursive inequality
\begin{align} 
\g(n) &\leq \sup_{x\geq n}\Bigl[
\big\{\big(\g(n-1)\mmin \g(x-n)\big)(1 + k\,q^{2(x-n)})
\big\}    \mmax k'\,\Bigr] \nonumber \\ 
      &\leq \sup_{n\leq x \leq 2n-1}
         \Bigl[\, (\g(x-n)\vee k'')(1 +
         k\,q^{2(x-n)}) \Bigr]\,.
\la{8d} 
\end{align}
It is quite simple now to conclude that 
$\g(n)$ is uniformly bounded. 
Indeed if $\tga(m):=(\g(m)\vee k'')$ then (\ref{8d}) tells us that
\be
\tga(m)\leq \sup_{1\leq \ell \leq m -1} \Big[
\tga(\ell)(1+kq^{2\ell})\Big]\,.
\la{9d}
\end{equation}
We then have a sequence $\ell_1<\ell_2<\cdots<\ell_s$, $s\leq m-1$ such
that 
$$
\tga(m)\leq \tga(1)\,\prod_{i=1}^s (1+kq^{2\ell_i})
$$ 
which is finite since $\tga(1)<\infty$ and
$\sum_i q^{2\ell_i} <\infty$. 
Thus $\tga(m)$ is uniformly bounded and so is $\g(n)$.   \qed




\section{Proof of Theorem \ref{transport}}
The setting in this section is as in (\ref{divide}). 
For notation convenience in what follows we will drop the subscripts
$\L,n$. We also use $\nu(\cdot\tc m)$ for $\nu(\cdot\tc N_{\L_1}= m)$.
If we apply Proposition \ref{glauber} to the function
$g(N_{\La_1})=\nu\big(f\,|N_{\La_1}\big)$ we get 
\be
\Var_\nu
\big(g(N_{\La_1})\big)\leq
k\big(n\wedge L^2\big) \sum_m p_n(m)\wedge p_n(m+1)
\left[g(m+1)-g(m)\right]^2
\end{equation}
where $p_n(m) = \nu\big(N_{\La_1}=m\big)$.\acapo
Therefore we need
to study the gradient $g(m+1)-g(m)$. For this purpose the main idea (very
roughly) is the following. 
\acapo 
Pick a configuration $\a$ such that
$N_{\L_1}(\a)=m+1$ and $N_\L(\a)=n$, choose two sites $x\in \L_1,\,y\in
\L_2$ such that $\a(x)=1, \,\a(y)=0$ and consider the exchanged
configuration $\h=\a^{xy}$. Clearly $N_{\L_1}(\h)=m$ and
$N_\L(\h)=n$. Using this kind of change of variables it is not difficult
to write an expression for the gradient $g(m+1)-g(m)$ in terms of
suitable spatial averages of $\nabla_{xy}f$ plus a covariance term
$\nu(f,F_{xy})$, where the latter originates from the action of the
change of variables on the probability measures. \acapo One possibility
to concretely implement this program is to write
$$
\nu\big(f\tc m+1\big) = 
\frac{1}{(m+1)(n-m-1)} 
\sum_{x\in \L_1, y\in\L_2}\nu\big(f\,\a_x(1-\a_y)\tc m+1\big)
$$
and to make the change of variables described above for each pair
$(x,y)$. This idea works just fine in the context of translation
invariant lattice gases \cite{CancMart}, but has some drawback in our
context due to the nature of the typical configurations of the measure
$\nu\big(\cdot \tc m+1\big)$.  As already shown, the
$m+1$ particles in $\L_1$ tend to fill the cylinder $\L_1$ up to a well
specified height and the same for $\L_2$. Without loss of generality we
can assume $m \geq n/2$ so that the resulting surface in $\L_1$
will stay higher than the surface in $\L_2$. Thus, if we don't want to
transform a typical configuration of $\nu\big(\cdot \tc
m+1\big)$ into an atypical one for $\nu\big(\cdot \tc m\big)$
via the exchange $T_{xy}$, we should only try to exchange the holes that
sit on the surface in $\L_2$ with the particles on the surface in
$\L_1$. In other words the above (deterministic) sum
$$
\sum_{x\in \L_1, y\in\L_2} \a_x(1-\a_y)
$$ 
should be replaced by a random variable 
$$
\sum_{x\in A,\, y\in B} \a_x(1-\a_y)
$$
where $A,B$ denote the two surfaces. Of course, for certain rare
configurations, the surfaces either do not exist or their density of
particles is far from its typical value. We are forced therefore to
split according to some criterium the contribution to the gradient
$g(m+1)-g(m)$ coming from typical and rare configurations and apply the
above reasoning only to the typical cases. The contribution coming from
the rare configurations should be estimated via moderate deviation bounds
for the measure $\nu\big(\cdot \tc m+1\big)$.\acapo
We will now make precise what we just said. In the rest of this section
we will always assume $m\ge \frac{n}{2}$.

\smallno
For any event $G\subset\Om_\La$ we write
\begin{align}
\bigl[\nu\big(f\,\tc m+1\big)& - 
\nu\big(f\,\tc m\big)\bigr]\nu(G\tc m+1)
= \nonumber \\  
\big[\nu\big(f\,\id_{G}\,\tc m+1\big) -
\nu\big(f\,\tc m\big)&\nu\big(G\,\tc m+1\big)\big]
- \nu\big(f,\id_{G}\,\tc m+1\big)\,.
\la{varg}
\end{align}
We then estimate
\begin{align}
\Big[\nu\big(f&\,\tc m+1\big) - \nu \big(f\,\tc m\big) \Big]^2
\leq 
2\frac{\nu\big(G^c\,\tc m+1\big)}
{\nu\big(G\,\tc m+1\big)}\Var_\nu\big(f\tc m+1\big)
\nonumber\\
&+ 2 \frac{1}
{\nu\big(G\,\tc m+1\big)^2}\left[\nu\big(f\id_{G}\,\tc m+1\big) - 
\nu\big(f\,\tc m\big)\nu\big(\id_{G}\,\tc m+1\big)\right]^2
\la{varg1}
\end{align}

\subsection{The typical events}
We will provide different definitions of the typical event $G$ according
to whether $L$ is ``large'' or ``small'' and whether we have ``many''
{\tt MP} or ``few'' {\tt FP} particles (see beginning of \S 3.5).

\medno
{\tt $L$ large}.

\noindent
We start with the {\tt MP} case.
\acapo 
Take $\l,\l'\in\bbR$ such that
\be
\mu_{\L_1}^\l(N_{\L_1})=m\,,\quad\mu_{\L_2}^{\l'}(N_{\L_2})=n-m\,.
\la{lambdas}
\end{equation}
Set 
\be
\ell_A=[\l\vee 0]\wedge (H-1)\,,\quad 
\ell_B=[\l'\vee 0]\wedge (H-1)
\la{k,l}
\end{equation} 
and define $A,B$ as the planar sections of $\L_1$ and $\L_2$ at height
$\ell_A$ and $\ell_B$ respectively. More precisely
\be
A=\La_1\cap\cA_{\ell_A}\qquad B=\La_2\cap\cA_{\ell_B}
\la{surface}
\end{equation}
Define the number of particles in $A$
and the number of holes in $B$
\be
N_A(\al)=\sum_{x\in A} \al_x\,,\quad V_B(\al)= \sum_{y\in B} (1-\al_y)\,,
\la{navb}
\end{equation}
and define $\wb N_A=\nu(N_A\tc m)$, $\wb V_B=\nu(V_B\tc m)$.

\begin{definition}
The event $G$ in the {\tt MP, $L$ large} case.
\acapo
We set  $G=G_A\cap G_B$, 
where
\be
G_A=\{|N_A(\al)-\wb N_A|\leq (n\wedge L^2)^{\frac{1}{2}+\gamma}\}\,,
\quad
G_B=\{|V_B(\al)-\wb V_B|\leq L^{1+2\gamma}\}\,.
\la{good}
\end{equation}
Here $\gamma$ is a small positive number, say $\gamma=.001$. 
\end{definition}

\noindent
Note that in
any case when $L$ is large $G$ implies
\be
N_A\geq \frac14 (n\wedge L^2)\,,\quad V_B \geq L^2\,.
\la{moda}
\end{equation}
\noindent
We turn to the case ({\tt FP}).  Here we do not fix two planar sections
but rather confine most of the particles in a cylinder with finite
height.  Let $h\in\bbN$, $1\leq h\leq H$ and define
\be
A = \big\{x\in\La_1:\;\ell_x\leq h-1\big\}\,,\quad 
B = \big\{y\in\La_2:\;\ell_y=0\}\,.
\la{gk1}
\end{equation}
\begin{definition}The event $G$ in the {\tt FP, $L$ large} case.\acapo
We set $G=G_A\cap G_B$, with
\be
G_A=\left\{
N_A(\al)\geq  m/2 \right\},\quad
G_B=\left\{|V_B(\al)-\wb V_B|\leq L^{1+2\gamma}\right\}\,.
\la{good3}
\end{equation}
\end{definition}

\noindent
Note that here too when $L$ is large $G$ implies
\be
N_A\geq \frac14 n\,,\quad V_B \geq L^2\,.
\la{modafp}
\end{equation}

\bigno
Finally we analyze the case of $L$ small. The construction of the sets
$A$ and $B$ is done exactly as before in the two cases {\tt MP} and {\tt
FP} but the definition of $G$ changes.

\begin{definition}The event $G$ in the {\tt $L$ small} case.
\acapo
We set  $G=G_A\cap G_B$, 
where
\be
G_A=\left\{N_A(\a) \ge 1\right\}\,,
\quad
G_B=\left\{V_B(\al)\ge 1\right\}\,.
\la{mpsmall}
\end{equation}
and $A$ and $B$ are as in (\ref{surface}) or as in (\ref{gk1}) depending
on whether we are in the {\tt MP} or the {\tt FP} case.
\end{definition}
 

\subsection{Bounds on the probability of the typical events}
In what follows we will provide some simple estimates on the probability
of the event $G^c$ in the various cases of few/many particles and large/small $L$.  
We begin by stating our bounds to be used when $L$ is large.
\begin{Le}
\la{moderate}
Assume {\tt (MP)}.  For any $q\in(0,1)$ there exist $k<\infty$ such that
\be
\nu\big(G^c\tc m\big)\leq k\exp({-k^{-1}L^{4\gamma}})\,.
\la{mod0}
\end{equation}
\end{Le}
\proof
Observe that
$$
\nu\big(G^c\tc m\big)\leq 
\nu_{\La_1,m}\big[G_A^c\big] +
\nu_{\La_2,n-m}\big[G_B^c\big]\,.
$$
We first prove
\be
\nu_{\La_1,m}\big[G_A^c\big]\leq k\exp({-k^{-1}L^{4\gamma}})\,.
\la{mod1}
\end{equation}
We write $A=A_1\cup A_2$ with $A_1=\Si_{L,H}\cap\cA_{\ell_A}$ and
$A_2 = \{(0,L,0) + \Si_{L,H}\}\cap\cA_{\ell_A}$. Letting
$$
G_i = \big\{ |N_{A_i}(\al)-\wb N_{A}/2|\leq \frac12 (n\wedge
L^2)^{1/2+\gamma} \big\},\quad i=1,2\,,
$$
we see that 
\be
\nu_{\La_1,m}\big[G_A^c\big]\leq 
\nu_{\La_1,m}\big[G_1^c\big] + \nu_{\La_1,m}\big[(G_2)^c\big]
\la{mod10}
\end{equation}
By Proposition \ref{canvar} we can estimate
(\ref{mod10}) with the help of the grand canonical distribution
$\mu^\lambda$ where $\l$ is given by (\ref{lambdas}). Thus 
\be
\nu_{\La_1,m}\big[G_A^c\big]\leq 
k\mu^{\lambda}\big[G_1^c\big] + 
k\mu^{\lambda}\big[(G_2)^c\big] 
= 2k \mu^{\lambda}\big[G_1^c\big]\,.
\la{mod2}
\end{equation}
Note that $\al_x$, $x\in A_1$ are i.i.d.\ random variables
under $\mu^{\lambda}$, with mean value $\rho_x=\mu^{\lambda}(\al_x)$.
Let us consider the case $n\leq L^2$ in detail. 
For the case $n\geq L^2$ simply replace $n$ by $L^2$
in the lines below. We have, for any $t\geq 0$
\be
\mul\big[G_1^c\big]\leq \exp{(-t\,n^{1/2+\gamma})}
\left[\exp{|A_1|\varphi(t)}+\exp{|A_1|\varphi(-t)}\right]\,,
\la{mod3}
\end{equation}
where
\be
\varphi(t)=\log{\mu_{\L_1}^\l\big(\exp{t\,(\al_x-\rho_x)}\big)}\,.
\la{mod4}
\end{equation}
Then $\varphi(0)=\varphi'(0)=0$ and $\varphi''(t)=\var^{\l_t}(\al_x)$,
where $\l_t=\l+t/(-2\log{q})$. Now, for any $|t|\leq 1$,
\be
\var_{\mu^{\l_t}}(\al_x)=\mu^{\l_t}
         \left[(\al_x-\mu^{\l_t}(\al_x))^2\right]
\leq \mu^{\l_t}\left[(\al_x-\mu^{\l}(\al_x))^2\right]
\leq {\rm e^2} \var_{\mu^{\l}}(\al_x)\,.
\la{mod5}
\end{equation}
Then 
\be
|\varphi(t)|\leq 5 \var_{\mu^{\l}}(\al_x) t^2\,,\quad |t|\leq 1\,.
\la{mod6}
\end{equation}
Using Lemma \ref{varnumber} and Lemma \ref{meannumber}
we have 
$$
|A_1|\var_{\mu^{\l}}(\al_x)=\var_{\mu^\l}(N_{A_1})\leq km\leq kn \,. 
$$
Therefore by (\ref{mod3}), choosing $t=O(n^{-1/2+\gamma})$ we obtain
\be
\mu^\l \big[G_1^c\big]\leq k\exp(-k^{-1}n^{2\gamma})\leq 
k'\exp(-{k'}^{-1}L^{4\gamma})
\la{mod7}
\end{equation}
The estimate for $V_B$ is obtained in a similar fashion. \qed

\bigskip
We turn to analyze the case of few particles. Again our estimate will be
meaningful only if $L$ is large enough.
\begin{Le} \la{how-good2} 
Assume {\tt (FP)}.  For any $q\in(0,1)$
there exist $k<\infty$, $h_0<\infty$ such that for all $m\in [\frac{n}{2},n]$
and $h\geq h_0$ we have
\be 
\nu\big(G^c\tc m\big)\leq k\big(
n^{-1}q^{2h}+\exp{(-k^{-1}L^{4\gamma})}\big)\,.  
\la{gk2} 
\end{equation} 
\end{Le} 

\proof 
Repeating the argument leading to (\ref{mod3}) and (\ref{mod6}),
choosing $t=O(L^{-1+2\gamma})$, we easily
obtain  
\be
\nu_{\La_2,n-m}\big[G_B^c\big]\leq k \exp{(-tL^{1+2\gamma})}
\exp{(knt^2)}\leq k'\exp{(-{k'}^{-1}L^{4\gamma})}\,.  
\la{gk111} 
\end{equation} 
Let $\bar A=\La_1\setminus A$ and write 
$$
\nu_{\La_1,m}\big[G_A^c\big]\leq \nu_{\La_1,m}\big[N_{\bar A}\geq
m/2\big]\,.  
$$ 
Dividing $\bar A$ in two parts $\bar A=A_1\cup A_2$ with
$A_1=\Si_{L,H}\cap \bar A$ and
$A_2=\{(0,L,0)+\Si_{L,H}\}\cap\bar A$, we may estimate 
$$
\nu_{\La_1,m}\big[N_{\bar A}\geq m/2\big] \leq
\nu_{\La_1,m}\big[N_{A_1}\geq m/4\big] + \nu_{\La_1,m}\big[N_{A_2}\geq
m/4\big]\,.  
$$ 
Then by Proposition \ref{canvar} it is sufficient to estimate
$\mu_{\L_1}^\l[N_{A_1}\geq m/4]$, where $\l$ is given by (\ref{lambdas}).
Since $m\leq n\leq \d L^{2}$ we have $\l \leq 0$ from Lemma
\ref{meannumber}.  Therefore 
\begin{align*} 
\mu_{\L_1}^\l(N_{A_1}) &=
L^2\sum_{j=h}^{H-1}\frac{q^{2(j-\l)}}{1+q^{2(j-\l)}} \leq
2q^{2h}L^2\sum_{j=0}^{H-1}\frac{q^{2(j-\l)}}{1+q^{2(j-\l)}}\\ & =
2q^{2h}\mu_{\L_1}^\l(N_{\La_1})=2q^{2h} m\,.  
\end{align*} 
We then estimate
$$ 
\mu_{\L_1}^\l\big[N_{A_1}\geq m/4\big]\leq \mu_{\L_1}^\l\big[
|N_{A_1}-\mu_{\L_1}^\l(N_{A_1})| \geq c m\big] \,, 
$$ 
with $c>0$, if
$h\geq h_0(q)$ for some $h_0(q)<\infty$. Then 
\begin{align} 
\mu_{\L_1}^\l[N_{A_1}\geq m/4] &\leq
\frac{1}{c^2m^2}\var_{\mu_{\L_1}^\l}(N_{A_1}) = 
\frac{L^2}{c^2m^2}\sum_{j=h}^H
\frac{q^{2(j-\l)}}{(1+q^{2(j-\l)})^2} \nonumber\\ & \leq
\frac{(1-q^2)^{-1}q^{2(h-\l)}L^2}{c^2m^2} \leq \frac{4q^{2h}}{c^2m}\,,
\la{gk10} 
\end{align} 
where in the last bound we use $$ q^{-2\l}\leq
\frac{4m}{L^2}(1-q^2)\,, $$ which follows from Lemma \ref{meannumber}.
Since $m\geq n/2$ (\ref{gk10}) and (\ref{gk111}) yield (\ref{gk2}). \qed

\bigskip

Finally we analyze the case $L$ small. Below the event $G$
will be that appearing in (\ref{mpsmall}) 
\begin{Le}\la{Lsmall}
\be
\nu(G\tc m) \geq c(q,L) > 0\,,
\la{aw1}
\end{equation}
with a constant $c(q,L)$ 
independent of the height $H$ of the cylinder.
\end{Le}
\proof  
Since 
\be
\nu(G_A\tc m) = 
\frac{\mu_{\L_1}^\l(G_A\cap\{N_{\L_1}=m\})}
     {\mu_{\L_1}^\l(\{N_{\L_1}=m\})}\,,
\la{aw2}
\end{equation}
the claim easily follows from a slight modification of the argument given
at the end of Theorem \ref{equiv} (packing all particles at the bottom
of the cylinder). The same can be done for the event $G_B$. \qed

\subsection{Bounding the gradient $\left[\nu\big(f\id_{G}\,\tc
m+1\big) - \nu\big(f\,\tc m\big)\nu\big({G}\,\tc
m+1\big)\right]^2$}
From Lemmas \ref{moderate}, \ref{how-good2} and \ref{Lsmall} we
see that, for any $\eps\in (0,1)$ the first term in the r.h.s.\ of
(\ref{varg1}) can be bounded from above by
\acapo
{\tt (i) $L$ large}.
\be
2\frac{\nu\big(G^c\,\tc m+1\big)}
{\nu\big(G\,\tc m+1\big)}\Var_\nu\big(f\tc m+1\big)
\le \frac{\eps}{n\wedge L^2}\Var_\nu\big(f\tc m+1\big)
\la{pezzo1}
\end{equation}
provided that $L$ and the constant $h$ in Lemma \ref{how-good2} are
large enough depending on $\eps$.
\acapo
{\tt (ii) $L$ small}.
\be
2\frac{\nu\big(G^c\,\tc m+1\big)}
{\nu\big(G\,\tc m+1\big)}\Var_\nu\big(f\tc m+1\big)
\le C\Var_\nu\big(f\tc m+1\big) 
\la{pezzo1bis}
\end{equation}
where $C=C(L)$ is some finite constant independent of $m$.

\bigno
We now turn our attention to the second term appearing in the r.h.s.\ of
(\ref{varg1}). As before, the factor $2 \nu\big(G\,\tc
m+1\big)^{-2}$ can be bounded from below by either $\frac{2}{(1-\eps)^2}$ or
by $C'(L)$ for a suitable constant $C'(L)$ according to whether $L$ is
large enough (depending on $\eps$) or it is small (i.e.\ smaller than some
$L_0$). \acapo We thus concentrate on the computation of the relevant
term
$$
\left[\nu\big(f\id_{G}\,\tc m+1\big) - 
\nu\big(f\,\tc m\big)\nu\big({G}\,\tc m+1\big)\right]^2\,.
$$
%
%
The following calculation holds irrespectively of which definition of
$G$ is adopted.
\acapo 
Defining
$$
\phi_{xy}(\al)=\frac{\al_x(1-\al_y)}{N_A(\al)V_B(\al)}\,,\quad
x\in A,\,y\in B\,,
$$
we may write
\be
\nu\big(f\id_{G}\,\tc m+1\big)=
\sumtwo{x\in A}{y\in B}\nu\big(f\id_{G}\phi_{xy}\,\tc m+1\big)\,.
\la{cd2}
\end{equation}
With the change of variables $\al\to T_{xy}\al$, (\ref{cd2}) becomes
\begin{align}
\nu\big(f\id_{G}\,\tc m+1\big)&=
\frac{p_n(m)}{p_n(m+1)}
\sumtwo{x\in A}{y\in B}q^{2(\ell_x-\ell_y)}\nu\big(T_{xy}\,[f
\id_{G}\phi_{xy}]\,\tc m\big)\nonumber\\
& = \si_m\sumtwo{x\in A}{y\in B}q^{2(\ell_x-\ell_y)}\nu\big(T_{xy}f
F_{xy}\,\tc m\big)\,,
\la{cd3}
\end{align}
where 
\be
\si_m=\frac{p_n(m)}{p_n(m+1)}\,,
\quad F_{xy}(\al)=\id_{G}(\al^{xy})\phi_{xy}(\al^{xy})\,.
\la{cd31}
\end{equation}
Subtracting and adding $f$ inside averages gives
\begin{equation}
\nu\big(f\id_{G}\,\tc m+1\big) =\si_m
\sumtwo{x\in A}{y\in B}q^{2(\ell_x-\ell_y)}\left\{
\nu\big(\grad_{xy}f
F_{xy}\,\tc m\big)
+ \nu\big(fF_{xy}\,\tc m\big)
\right\}\,.
\la{cd4}
\end{equation}
When $f={\bf 1}$ we see that 
\be
\nu\big({G}\,\tc m+1\big)
= \si_m
\sumtwo{x\in A}{y\in B}q^{2(\ell_x-\ell_y)}
\nu\big(F_{xy}\,\tc m\big)\,.
\la{cd41}
\end{equation}
Therefore, by subtracting
$\nu\big(f\,\tc m\big)\nu\big({G}\,\tc m+1\big)$ the last
term in (\ref{cd4}) becomes a covariance
\begin{align}
\nu&\big(f\id_{G}\,\tc m+1\big) - 
\nu\big(f\,\tc m\big)\nu\big({G}\,\tc m+1\big)\nonumber\\
&= \si_m
\sumtwo{x\in A}{y\in B}q^{2(\ell_x-\ell_y)}
\left\{\nu\big([\grad_{xy}f]\,
F_{xy}\,\tc m\big)
+ \nu\big(f,F_{xy}\,\tc m\big)\right\}\,.
\la{cd5}
\end{align}
We then estimate the square of the l.h.s.\ of (\ref{cd5}) by
\be
\left[\nu\big(f\id_{G}\,\tc m+1\big) - 
\nu\big(f\,\tc m\big)\nu\big({G}\,\tc m+1\big)\right]^2
\leq \,\bbI_1\,+\,\bbI_2\,,
\la{cd6}
\end{equation}
with 
\be
\bbI_1=2\Big\{\si_m
\sumtwo{x\in A}{y\in B}q^{2(\ell_x-\ell_y)}
\nu\big(\grad_{xy}f
F_{xy}\,\tc m\big)\Big\}^2\,,
\la{cd7}
\end{equation}
and 
\be
\bbI_2=2\Big\{\si_m
\sumtwo{x\in A}{y\in B}q^{2(\ell_x-\ell_y)}
\nu\big(f,
F_{xy}\,\tc m\big)\Big\}^2\,.
\la{cd8}
\end{equation}
%

\acapo
\medno
\centerline{\tt Estimate of $\bbI_1$}
Using (\ref{cd41}), the
non-negativity of $F_{xy}$ and the Schwarz' inequality we obtain
\begin{align}
\bbI_1 &\leq 2\si_m
\sumtwo{x\in A}{y\in B}q^{2(\ell_x-\ell_y)}
\left\{\frac{\nu
\left(\grad_{xy}fF_{xy}\,\tc m\right)}{\nu\big(F_{xy}\,\tc m\big)}
\right\}^2
\nu\big(F_{xy}\,\tc m\big)\nonumber\\
&\leq 2\si_m
\sumtwo{x\in A}{y\in B}q^{2(\ell_x-\ell_y)}
\nu\big((\grad_{xy}f)^2F_{xy}\,\tc m\big)
\la{cd9}
\end{align}
Next we observe that, by the definition of the event $G$, using
(\ref{moda}) and (\ref{modafp}) we have
\be
F_{xy}(\al)=
\id_{G}(\al^{xy})\frac{\al_y(1-\al_x)}{N_A(\al^{xy})V_B(\al^{xy})}
\leq \begin{cases}
4(n\wedge L^2)^{-1}L^{-2}\al_y(1-\al_x) & {\tt (MP)} \text{ $L$ large}\\
4n^{-1}L^{-2}\al_y(1-\al_x) & {\tt (FP)} \text{ $L$ large}\\
\al_y(1-\al_x) & \text{ $L$ small}
\end{cases}
\la{mp0}
\end{equation}
Therefore in both cases ({\tt MP}), ({\tt FP}) 
\be
\bbI_1 \leq 4\si_m L^{-2}(n\wedge L^2)^{-1}
\sumtwo{x\in A}{y\in B}q^{2(\ell_x-\ell_y)}
\nu\big((\grad_{xy}f)^2\al_y(1-\al_x)\,\tc m\big)
\la{mp1}
\end{equation}
if $L$ is large, while 
\be
\bbI_1 \leq 
\si_m\sumtwo{x\in A}{y\in B}q^{2(\ell_x-\ell_y)}
\nu\big((\grad_{xy}f)^2\al_y(1-\al_x)\,\tc m\big)
\la{mp1bis}
\end{equation}
if $L$ is small.
\smallno
We can finally apply Proposition \ref{I1} to obtain
\be
\bbI_1\leq \si_m\begin{cases}
\frac{C}{n\,\wedge\, L^2}
\Big\{ L^2 \sum_{b\in\cO_\La}
\nu\big((\grad_{b}f)^2\,\tc m\big)
+ \sum_{b\in\cV_\La}
\nu\big((\grad_{b}f)^2\,\tc m\big)\Big\} & \text{ if $L$ is
large} \\
\\
\Big\{ L^2 \sum_{b\in\cO_\La}
\nu\big((\grad_{b}f)^2\,\tc m\big)
+ \sum_{b\in\cV_\La}
\nu\big((\grad_{b}f)^2\,\tc m\big)\Big\} & \text{ if $L$ is
small} 
\end{cases}
\la{I1final} 
\end{equation}
where 
$C$ is a suitable constant depending on $q$ and $h$
($h$ is the constant in Lemma \ref{how-good2}).
%
%
%
%

\acapo
\medno
\centerline{\tt Estimate of $\bbI_2$}
Recall the definition of $\bbI_2$ given in (\ref{cd8}). It is quite
clear from (\ref{mp0}), Lemma \ref{sinm} and the Schwartz
inequality that
\be
\bbI_2 \le k\,L^8 \var_\nu(f\tc m) 
\la{I2Lsmall}
\end{equation}
where $k=k(q)$. Such a bound will turn out useful when $L$ is
``small''. The case $L$ large is more involved and requires a more
subtle analysis. We start with the case ({\tt MP}).
\begin{Le}
\la{corre}
For every $\eps>0$ and $q\in(0,1)$ there exist finite constants $C_\eps$
and $L_0$ such that for any $L\ge L_0$, $H,n$ satisfying {\tt (MP)} the
following estimate holds
\be
\bbI_2\leq (n\wedge L^2)^{-1}\left\{C_\eps\,\Dir_\nu(f,f\tc m) + 
\eps\,\Var_\nu(f\tc m)\right\}\,.
\la{corre1}
\end{equation}
\end{Le}
\proof
From (\ref{cd8}) and Lemma \ref{sinm} we have a first estimate
\be
\bbI_2\leq k\Big\{
\sumtwo{x\in A}{y\in B}\nu\big(f,F_{xy}\,\tc m\big)\Big\}^2\,.
\la{corre2}
\end{equation}
Observe that 
\be
\sumtwo{x\in A}{y\in B} F_{xy}=\frac{V_AN_B\id_{\wt G}}{(N_A+1)(V_B+1)}
\la{corre3}
\end{equation}
where $\wt G=\wt G_A\cap\wt G_B$ with 
$$
\wt G_A=\{|N_A(\al)+1-\wb N_A|\leq (n\wedge L^2)^{1/2+\gamma}\}
$$ and
$$
\wt G_B=\{|V_B(\al)+1-\wb V_B|\leq L^{1 + 2\gamma}\}\,.
$$
As in Lemma \ref{moderate} we have the bounds 
\be
\nu({\wt G_A}^c\tc m)\leq k\exp{(-k^{-1}L^{4\gamma})}\,,\quad
\nu({\wt G_B}^c\tc m)\leq k\exp{(-k^{-1}L^{4\gamma})}\,.
\la{corre5}
\end{equation}
Writing
$$
F_A= \frac{V_A}{N_A+1}\id_{\wt G_A}\,,\quad F_B= \frac{N_B}{V_B+1}\id_{\wt G_B}
\,,
$$ 
(\ref{corre2}) says that
\be
\bbI_2\leq k\, \nu\big(f,F_AF_B\big|m\big)^2 
\la{corre6}
\end{equation}
We write $\nu(\cdot\tc m) = \nu_1\otimes\nu_2$ where
$\nu_1=\nu_{\La_1,m}$ and $\nu_2=\nu_{\La_2,n-m}$ and use the decomposition
\be
\nu\big(f,F_AF_B\tc m\big)
=  \nu_2(F_B) \nu\big(f,F_A\tc m\big)
+ \nu\big(F_A\nu_2\big(f,F_B\big)\tc m\big)\,.
\la{corre7}
\end{equation}
We start by 
estimating $\nu\big(f,F_A\tc m\big)^2$.
Defining 
$$
\rho_A=\frac{N_A}{|A|}, \quad \wb \rho_A=\frac{\wb N_A}{|A|}
$$
we may write
\be
F_A = \frac{\id_{\wt G_A}}{\rho_A} - \id_{\wt G_A}
+ \id_{\wt G_A}\left[1-\frac1{\rho_A}\right]\frac1{N_A+1}\,.
\la{corre8}
\end{equation}
For the second term in the right side of (\ref{corre8}) one can use 
(\ref{corre5}). For the third term, recalling (\ref{moda}), 
one has an upper bound of
order $kL^{-2}$. 
Therefore apart from the first term the rest contributes 
at most $kL^{-4}\Var_\nu(f\tc m)$
to the upper bound on $\nu\big(f,F_A\tc m\big)^2$. 
The first term in (\ref{corre8}) is handled as follows.
We expand
\be
\frac{\id_{\wt G_A}}{\rho_A} = \frac{\id_{\wt G_A}}{\wb\rho_A} 
\left(2-\frac{\rho_A}{\wb\rho_A} + \cR_A\right)
\la{corre9}
\end{equation}
where, on $\wt G_A$, 
\be
|\cR_A|\leq k\left(\frac{\rho_A}{\wb\rho_A}-1\right)^2\leq 
k (n\wedge L^2)^{-1+2\gamma}\,.
\la{corre10}
\end{equation}
In view of (\ref{corre10}) and using again (\ref{corre5}) to depress the term
proportional to $\id_{\wt G_A}$ we have obtained
\be
\nu\big(f,F_A\tc m\big)^2\leq k({\wb\rho_A}^2{\wb N_A}^2)^{-1}
\,\nu\big(f,N_A\tc m\big)^2  + k L^{-4+8\gamma} 
\Var_\nu(f\tc m)\,.
\la{corre11}
\end{equation}
An application of Proposition \ref{twoblock} together with the bound
\hbox{ $({\wb\rho_A}^2{\wb N_A}^2)^{-1}\leq kL^{4}(n\wedge L^2)^{-4}$} 
yields the estimate
\be
\nu\big(f,F_A\tc m\big)^2\leq L^{4}(n\wedge L^2)^{-3}
\left\{C\,\Dir_\nu(f,f\tc m) + 
\eps\Var_\nu(f\tc m)\right\}\,.
\la{corre12}
\end{equation}
Using $F_B\leq (n\wedge L^2)L^{-2}$, the first term in (\ref{corre7}) 
can be finally estimated by
\be
\nu_2(F_B)^2 \nu\big(f,F_A\tc m\big)^2\leq
(n\wedge L^2)^{-1}\left\{C\,\Dir_\nu(f,f\tc m) + 
\eps\Var_\nu(f\tc m)\right\}\,.
\la{corre13}
\end{equation}
We turn to the second term in (\ref{corre7}). Repeating the arguments
given above and using $|V_B-\wb V_B|\leq L^{1+2\gamma}$ we obtain the
upper bound
\be
\nu\big(\nu_2\big(f,F_B\big)^2\tc m\big)\leq 
kL^{-4}
\nu\big(\nu_2\big(f,N_B\big)^2\tc m\big)  
+ kL^{-4 + 8\gamma} \Var_\nu(f\tc m)\,.
\la{corre14}
\end{equation}
By Proposition \ref{twoblock}
\be
\nu\big(\nu_2\big(f,F_B)^2\tc m\big)\leq 
(n\wedge L^2)L^{-4}\left\{C\,\Dir_\nu(f,f\tc m) + 
\eps\Var_\nu(f\tc m)\right\}\,.
\la{corre15}
\end{equation}
Recalling that $F_A\leq L^2(n\wedge L^2)^{-1}$ we can estimate the
square of the second term in (\ref{corre7}) by
\be
\nu(F_A^2\tc m) \nu\big(\nu_2\big(f,F_B)^2\tc m\big)\leq 
(n\wedge L^2)^{-1}\left\{C\,\Dir_\nu(f,f\tc m) + 
\eps\Var_\nu(f\tc m)\right\}\,.
\la{corre13bis}
\end{equation}
\qed

We now turn to estimate $\bbI_2$ in the case ({\tt FP}).
Recall that here $A$ is the cylinder with height $h$, see (\ref{gk1}).
%

\begin{Le}
\la{correfp}
For every $\eps>0$, $q\in(0,1)$, there exists $\d_0(\eps,q)>0$ and
finite constants $C_\eps$, 
$L_\eps$ and $h_\eps$ such that for any $L\ge L_\eps$, $H,n$
satisfying {\tt (FP)} with $\d\leq \d_0$, 
any $m\ge \frac{n}{2}$ and $h\geq h_\eps$
\be
\bbI_2\leq n^{-1}
\left\{C_\eps\,\cE_\nu(f,f\tc m)+\eps\, \Var_\nu(f\tc m)\right\}\,.
\la{correfp1}
\end{equation}
\end{Le}
\proof
Define
$$
F_A= 
\frac{\id_{\wt G_A}}{N_A+1}
\sum_{j=0}^{h-1}q^{2j}V_j\,,\quad F_B= \frac{N_B}{V_B+1}\id_{\wt G_B}
\,,
$$ 
with 
$$
V_j=\sum_{x\in A_j}(1-\al_x)\,,\quad A_j=\{x\in A: \ell_x=j\}\,,
$$
and $\wt G_A=\{N_A+1\geq m/2\}$, 
$\wt G_B=\{|V_B+1-\wb V_B|\leq L^{1+2\gamma}\}$. 
Then as in (\ref{corre2}) we have
\be
\bbI_2\leq k\, \nu\big(f,F_AF_B\tc m\big)^2 \,.
\la{correfp2}
\end{equation}
and we decompose as in (\ref{corre7}).
Let us first estimate 
\begin{align}
\nu(f,F_A\tc m)^2&\leq \var_\nu(f\tc m)\var_\nu(F_A\tc m)\nonumber\\
&=\var_\nu(f\tc m)\left[
\nu\big(\var_\nu(F_A \tc N_A)\tc m\big) + \var_\nu\big(\nu(F_A \tc N_A)\tc m\big)
\right]\,.
\la{correfp3}
\end{align}
Observe that 
\begin{align}
\var_\nu(F_A \tc N_A)&=\frac{\id_{\wt
G_A}}{(N_A+1)^2}\var_\nu\big(\sum_jq^{2j}V_j \tc N_A\big) 
\nonumber\\
&\leq kn^{-2}\sum_j q^{2j} \var_\nu\big(V_j\tc N_A\big)
\leq  k n^{-1}\,,
\la{correfp4}
\end{align}
where we used the fact that $N_A+1\geq m/2\geq n/4$ and 
that $\var_\nu (V_j \tc N_A)= \var_\nu (N_{A_j} \tc N_A) \leq k n$. The latter estimate
can be derived as usual from Proposition 
\ref{canvar} and Lemma \ref{varnumber}.
For the second term in (\ref{correfp3}) we claim that
\be
\var_\nu\big(\nu(F_A \tc N_A)\tc m\big)\leq k L^4 n^{-3} q^{2h}\,.
\la{correfp5}
\end{equation}
Set 
$$
\varphi_j(N_A)=\frac{\nu(V_j \tc N_A)}{N_A+1}
\,,
$$
so that
\be
\var_\nu\big(\nu(F_A \tc N_A)\tc m\big)\leq k \sum_j q^{2j}
\var_\nu(\varphi_j(N_A)\id_{\wt G_A}\tc m)\,.
\la{correfp6}
\end{equation}
We have
\begin{align}
\var_\nu(\varphi_j(N_A)\id_{\wt G_A}\tc m)&=
\sum_{\ell,\ell'} \nu(N_A=\ell\tc m)\nu(N_A=\ell'\tc m)\,\times\,\nonumber\\
\;&\times\,
\big(\varphi_j(\ell)-\varphi_j(\ell')\big)^2
\big[\id_{\wt G_A}(\ell)\id_{\wt G_A}(\ell')
+ 2\id_{\wt G_A}(\ell)\id_{\wt G_A^c}(\ell') \big]\,.
\la{correfp7}
\end{align}
Using $N_A\in [n/4,n]$, $V_j\leq kL^2$ 
and $|\nu(V_j \tc N_A=\ell)-\nu(V_j \tc N_A=\ell')|\leq n$,
(\ref{correfp7}) implies
\begin{equation}
\var_\nu(\varphi_j(N_A)\tc m)\leq k + kL^4n^{-4}\var_\nu(N_A\tc m) + 
kL^4n^{-2}\nu({\wt G_A^c}\tc m)\,.
\la{correfp8}
\end{equation}
From the equivalence of ensembles and Lemma \ref{varnumber} we
have
$$
\var_\nu(N_A\tc m)\leq k m q^{2h} \leq k n q^{2h}\,.
$$
Moreover, by Lemma \ref{how-good2} we know that
$\nu({\wt G_A^c}\tc m)\leq kn^{-1}q^{2h}$. Thus (\ref{correfp8})
combined with (\ref{correfp6}) yields the claim (\ref{correfp5}). 
Going back to (\ref{correfp3}) and recalling that 
$F_B\leq k n L^{-2}$ we have the estimate
\be
\nu_2(F_B)^2 \nu\big(f,F_A\tc m\big)^2\leq
k\big(q^{2h} n^{-1} + nL^{-4}\big) \var_\nu (f\tc m)\,.
\la{correfp9}
\end{equation}
Recall that $n\leq \d L^2$ so that $nL^{-4}\leq \d n^{-1}$ and we have to choose $\d$ small 
depending on $\eps$. 
We now estimate the term $\nu\big(\nu_2\big(f,F_B\big)^2\tc m\big)$
in (\ref{corre7}). As in (\ref{corre15})
we have
\be
\nu\big(\nu_2\big(f,F_B)^2\tc m\big)\leq 
n L^{-4}\left\{C_\eps\,\Dir_\nu(f,f\tc m) + 
\eps\Var_\nu(f\tc m)\right\}\,.
\la{correfp15}
\end{equation}
At this point the bound $F_A\leq kn^{-1}L^2$ gives 
\be
\nu(F_A^2\tc m)\nu\big(\nu_2\big(f,F_B)^2\tc m\big)\leq
n^{-1}\left\{C_\eps\,\Dir_\nu(f,f\tc m) + 
\eps\Var_\nu(f\tc m)\right\}\,.
\la{correfp16}
\end{equation}
Choosing $h$ sufficiently large in (\ref{correfp9}) and combining with (\ref{correfp16}) the proof
of (\ref{correfp1}) is complete. \qed

\bigno

\subsection{The proof of the theorem completed}

{\tt (i) $L$ large.}
\acapo
From the estimate of Proposition
\ref{glauber}applied to 
$g(N_{\L_1}):=\nu(f\tc N_{\L_1})$, (\ref{varg1}), 
the bound (\ref{pezzo1}) and (\ref{cd6}), we see that
\begin{gather}
\Var_\nu \big(\nu(f\tc N_{\La_1})\big)
\leq \nonumber\\
\eps \Var_\nu\big(f\big) 
+ k (n\wedge L^2)\sum_m p_n(m)\wedge p_n(m+1)
\big[ \bbI_1 + \bbI_2\big]\,,
\la{tr1}
\end{gather}
provided that $L$ is large enough depending on $q,\eps$. 
Thanks to (\ref{I1final})
\begin{align}
(n\wedge L^2)&\sum_m p_n(m)\wedge p_n(m+1) \,\bbI_1\nonumber\\
&\leq k \sum_m p_n(m) 
 \Big\{L^2 \sum_{b\in\cO_\La}
\nu\big((\grad_{b}f)^2\,|m\big)
+ \sum_{b\in\cV_\La}
\nu\big((\grad_{b}f)^2\,|m\big)\Big\}\nonumber\\
& = k\left\{L^2 \Dir_\nu^{\cO_\L}(f,f) + \Dir_\nu^{\cV_\L}(f,f)\right\} 
\la{tr2}
\end{align}
On the other hand the estimates on $\bbI_2$ given in Lemma
\ref{corre} and \ref{correfp} yield
\be
(n\wedge L^2)\sum_m p_n(m)\wedge p_n(m+1)\, \bbI_2\leq
C_\eps \,\Dir_\nu(f,f) + \eps \var_\nu(f) 
\,.
\la{tr3}
\end{equation}
for any $\eps >0$ and a suitable constant $C_\eps$ independent of $L$. 
In conclusion, for
any $\eps >0$ and $q\in (0,1)$ we can choose $L_0=L_0(\eps,q)$ such
that, by combining together (\ref{tr2}) and (\ref{tr3}),
the r.h.s.\ of (\ref{tr1}) is bounded from above by
\be
k\left\{L^2 \Dir_\nu^{\cO_\L}(f,f) + \Dir_\nu^{\cV_\L}(f,f)\right\}  +
C_\eps\, \Dir_\nu(f,f) + \eps \var_\nu(f) 
\end{equation}
for any $L\ge L_0$. 

\medno
{\tt (ii) $L$ small.}
\acapo
Using Proposition \ref{glauber} together with (\ref{pezzo1bis}),
and (\ref{cd6}), we see that
\begin{align}
\Var_\nu\left(\nu\big(f\,|N_{\La_1}\big)\right)
&\leq \nonumber\\ 
C 
\Big\{\sum_m
p_n(m)\wedge p_n(m+1)\big[ \bbI_1 + \bbI_2\big]
& + 
\Var_\nu(f)\,\Big\} 
\la{small1}
\end{align}
for a suitable constant $C= C(L,q)$. 
\acapo 
It is enough to use at this point (\ref{I1final}) together with
the rough estimate (\ref{I2Lsmall}) to get the sought bound.
\qed




\section{Proof of the upper bound in Theorem \ref{thetheorem} and of
Theorem \ref{teorema1}} 

In this final section we prove the upper bound on the spectral
gap of the generator $\cL_{\Si_{H,L},n}$ and the bound on the spectral
projection.


\subsection{Proof of (\ref{bound1}).} 
Consider the cylinder $\L:=\Si_{L,H}$ which has the square
$Q_L$ (containing $L^2$ sites) as 
basis. A generic point of $Q_L$ will be denoted by $z$
and $N_z$ stands for the number of particles
in the stick going through $z$, 
$$
N_z(\a)=\sum_{x\in\Si_{z,H}}\a_x\,.
$$
Given a smooth function $\varphi:[0,1]^2\to \bbR$,
we define $f_\varphi:\Om_\L\to\bbR$ by
\be
f_\varphi(\a)=\sum_{z\in Q_L} \varphi_L(z) N_z(\a)\,,
\la{upb2}
\end{equation}
where $\varphi_L$ denotes the rescaled profile
\be
\varphi_L(z)=\varphi(z/L)\,,\quad z\in Q_L\,.
\la{upb1}
\end{equation}
We will use the notation
$$
e(\varphi)=\int_{[0,1]^2}\|\grad\varphi(u)\|^2\dd u\,,\quad\quad
\|\grad\varphi(u)\|^2:=(\partial_{u_1}\varphi(u))^2 + 
(\partial_{u_2}\varphi(u))^2 \,.
$$
The upper bound in Theorem \ref{thetheorem} is obtained as follows.
\begin{Pro}
\la{upb}
For every $q\in(0,1)$, there exists $k=k(q)<\infty$ such that
the following holds.
For any smooth function $\varphi:[0,1]^2\to\bbR$ satisfying 
$\int \varphi(u)\dd u = 0$ and $\int\varphi(u)^2\dd u = 1$
%
there exists $L_0$ such that 
for any $L\geq L_0$, $H\geq 1$ and $n=1,\dots,HL^2-1$ one has
\be
\cE_\nu(f_\varphi,f_\varphi)
\leq k\,e(\varphi) \,L^{-2} \,\var_\nu(f_\varphi)\,,
\la{upb4}
\end{equation}
\end{Pro}
\proof
Observing that $N_z$, $z\in Q_L$ are identically distributed under $\nu$
we easily see that
\be
\nu(N_z,N_{z'})=-\frac{\si_\nu^2}{L^2-1} \,,\quad z\neq z'\,,
\la{upb5}
\end{equation}
where $\si_\nu^2:=\nu(N_z,N_z)$ is the variance of
the number of particles in a single stick. 
Thus
\begin{align}
\Var_\nu(f_\varphi)&=\si_\nu^2\sum_{z\in Q_L}\varphi_L(z)^2 -
\frac{\si_\nu^2}{L^2-1}\sumtwo{z,z'\in Q_L:}{z\neq z'}\varphi_L(z) 
\varphi_L(z')\nonumber\\
&\geq \si_\nu^2\sum_{z\in Q_L}\varphi_L(z)^2
- \frac{\si_\nu^2}{L^2-1}\left(\sum_{z\in Q_L}\varphi_L(z)\right)^2 \,.
\la{upb6}
\end{align}
Since $\int \varphi(u)\dd u = 0$ and $\int\varphi(u)^2\dd u = 1$, from 
Riemann integration we conclude that there exists a finite $L_0$ such that for any $L>L_0$
\be
\Var_\nu(f_\varphi)\geq \frac{\si_\nu^2}2 L^2\,.
\la{upb7}
\end{equation}
Let us now estimate the Dirichlet form.
In view of (\ref{upb7}) all we have to prove is
\be
\Dir_\nu(f_\varphi,f_\varphi)\leq k \,e(\varphi)\,\si_\nu^2 \,.
\la{upb8}
\end{equation}
Consider a bond $(x,y)=b\in\cO_\L$. 
Clearly 
$\grad_{xy} f_\varphi=0$ if $x,y$ belong to the same stick since an exchange
between $x$ and $y$ does not change the number of particles in any stick. 
In particular, only horizontal bonds $b\in\cO_\L$ contribute to $\cE_\nu(f,f)$. 
Take $z,z'\in Q_L$ such that 
$x\in\Si_{z,H}$, $y\in\Si_{z',H}$ and $b=(x,y)\in\cO_\L$. 
One has
\be
\grad_{xy}f_\varphi(\a)=(\a_y-\a_x)(\varphi_L(z)-\varphi_L(z'))
\la{upb9}
\end{equation}
and since $\|z-z'\|_1=1$,
$$
|\varphi_L(z)-\varphi_L(z')|\leq 2L^{-1} \|\grad\varphi(\tz/L)\| + O(L^{-2})\,.
$$
From (\ref{upb9}) we obtain,
\be
\Dir_\nu(f_\varphi,f_\varphi)\leq 2L^{-2}\sum_{z\in Q_L}
\big(\|\grad\varphi(\tz/L)\|^2 + O(L^{-2})\big)\, \cC(\nu,z) \,,
\la{upb10}
\end{equation}
with
$$
\cC(\nu,z) := \sum_{x\in\Si_{z,H}}\sumtwo{y\notin\Si_{z,H}:}{\|x-y\|_1=1}
\nu\big((\a_x-\a_y)^2\big)\,.
$$
Since
$$
L^{-2}\sum_{z\in Q_L}
\big(\|\grad\varphi(\tz/L)\|^2 + O(L^{-2})\big) 
\to \,e(\varphi)\,,
\quad\quad L\to\infty\,,
$$
the claim (\ref{upb8}) is proven once we show that there exists $k<\infty$
such that for any $z\in Q_L$ 
\be
\cC(\nu,z) 
\leq k \si_\nu^2\,.
\la{upb11}
\end{equation}
We start the proof of (\ref{upb11}) by estimating 
with the help of Proposition \ref{canvar}:
\be
\nu\big((\a_x-\a_y)^2\big) \leq k \mul\big((\a_x-\a_y)^2\big)\,,
\la{upb11.5}
\end{equation}
with $\mul$ the grand canonical measure corresponding to $n$ particles.
We observe that, since $x$ and $y$ are at the same height
$$
\mul\big((\a_x-\a_y)^2\big) = \mul(\a_x)(1-\mul(\a_x))
+\mul(\a_y)(1-\mul(\a_y)) = 
2\var_{\mul}(\al_x)
$$
For every $x\in\Si_z$ there are at most $4$ horizontal neighbours $y\notin\Si_z$ 
so that $\cC(\nu,z)\leq 8\,\si^2(\l)$, with $\si^2(\l):=\var_{\mul}(N_z)$.
The rest of the proof is now concerned with the estimate
\be
\si^2(\l)\leq k \si_{\nu}^2(n)
\la{upb14}
\end{equation}
with a constant $k$ only depending on $q$. 
Once (\ref{upb14}) is established we obtain (\ref{upb11})
and the proposition follows.

\smallskip

Below we restrict to the case $n\leq HL^2/2$, which is no loss 
of generality in view of particle-hole duality.
From (\ref{upb5}) we have
\be
\si_\nu^2 = \frac{L^2-1}{2L^2}\nu\big((N_z-N_{z'})^2\big)\,,\quad z\neq z'\,.
\la{upb16}
\end{equation}
For any integer 
$m\geq -1$ consider the event $E_{z,m}$ that the stick $\Si_{z,H}$ is
filled with particles up to level $m$ and is empty above level $m$. 
More precisely if 
$x_0=z,x_1,\dots,x_{H-1}$ are the sites of $\Si_{z,H}$
with $\ell_{x_i}=i$ we define
$$
E_{z,m}=\{\a_{x_0}=\dots=\a_{x_m}=1\,,\;\a_{m+1}=\dots=\a_{x_{H-1}}=0\}\,,\quad
E_{z,-1}=\{N_z=0\}\,.
$$
For any integer $0\leq m\leq H-1$ we have the bound
\be
\nu\big((N_z-N_{z'})^2\big) \geq \nu\big(E_{z,m}\cap E_{z',m-1})\big)\,.
\la{upb17}
\end{equation}
The right hand side above should be maximal around $m=[n/L^2]$.
Indeed, simple computations as in 
Lemma \ref{meannumber} show that there exists $\delta =\delta(q)>0$ such
that uniformly in the height $H$ one has
\be
\mul(E_{z,[n/L^2]})\geq \delta q^{-2(\l\wedge 0)}\,, \quad 
\mul(E_{z,[n/L^2]-1})\geq \delta\,.
\la{upb18}
\end{equation}
Therefore using Theorem \ref{equiv} we have
\begin{align}
\nu\big(E_{z,[n/L^2]}\cap E_{z',[n/L^2]-1}\big) 
&\geq \mul(E_{z,[n/L^2]})\mul(E_{z,[n/L^2]-1}) - 
kL^{-2} \nonumber\\
&\geq \delta^2 q^{-2(\l\wedge 0)} - kL^{-2}
\geq k^{-1}\delta^2 \si^2(\l) - kL^{-2}\,,
\la{upb19}
\end{align}
with the last inequality coming from Lemma \ref{varnumber}.
But we know (Remark \ref{vnvsn}) that 
$\si^2(\l) \geq k^{-1}
(1\wedge \frac n{L^2})$, thus (\ref{upb16}), (\ref{upb17})
and (\ref{upb19}) imply
that there exist finite 
constants $L_0,N_0$ and $k$ only depending on $q$
such that (\ref{upb14}) holds whenever $L\geq L_0$ and $n\geq N_0$. 

\smallskip

It remains to treat the case $n < N_0$. 
It will suffice to show 
\be
\si_\nu^2 \geq \frac{n}{kL^2}\,.
\la{upbclaim}
\end{equation}
We write
\begin{align*}
\nu\big((N_z-N_{z'})^2\big) &\geq \nu \big(N_z=1,N_{z'}=0\big)
\nonumber\\
&\geq \nu \big(N_z=1,N_{z'}=0\tc N_w\leq 1\,,\;\forall w\in Q_L\big)
\nu\big(N_w\leq 1\,,\;\forall w\in Q_L\big) \nonumber\\
& \geq \nu \big(N_z=1,N_{z'}=0\tc N_w\leq 1\,,\;\forall w\in Q_L\big)
\nu\big(\sum_{w\in Q_L}\a_w=n\big)
\la{upb20}
\end{align*}
But 
\begin{align*}
\nu\Big(\sum_{w\in Q_L}\a_w=n\Big) &= 
\frac{\mul(\sum_{w\in Q_L}\a_w=n\,,\;N_{\L\setminus Q_L}=0)}
{\mul(N_{\L}=n)}\\
&\geq \mul\Big(\sum_{w\in Q_L}\a_w=n\,,\;N_{\L\setminus Q_L}=0\Big) \,,
\end{align*}
and the latter is bounded away from $0$ uniformly 
as in the proof of Theorem \ref{equiv} (see (\ref{pois})).  
On the other hand
$$
\nu \big(N_z=1,N_{z'}=0\tc N_w\leq 1\,,\;\forall w\in Q_L\big)
= \frac{\binom{L^2-2}{n-1}}{\binom{L^2}{n}} = \frac{n(L^2-n)}{L^2(L^2-1)}
\geq \frac{n}{2L^2}\,,
$$
as soon as $L^2\geq 2N_0$. This yields the desired bound (\ref{upbclaim}).
\qed

\begin{remark}\la{excite}
The above proposition allows to produce low-lying excitations
which are localized in a sub-cylinder 
$\Si_{R,H}\subset\Si_{L,H}$ with $R\leq L$, much in the spirit of \cite{BCNS}. 
Indeed, one can 
always choose a function $\varphi$ supported on $[0,R/L]$ 
with $\int \varphi^2 = 1$ and $e(\varphi) = O(R^{-2}L^2)$
and the resulting states $f_\varphi$ have energy $O(R^{-2})$. 
\end{remark}

\subsection{Proof of Theorem \ref{teorema1}} 
For simplicity we prove the result for the generator $\cL_{\L,n}$ instead
of $\cG_{\L,n}$, but the argument applies essentially without modifications
to the original setting of Theorem \ref{teorema1}. 

\smallskip

We follow quite closely the
proof of an analogous result for translation invariant lattice gases
(see Theorem 2.4 in \cite{CancMart}). The main idea is
to establish the following inequality
\be
\nu(g,f)^2 \leq 
k_\eps \big\{\,\ell^{\eps} \,\cE_\nu(g,g) +
\ell^{-2}\nu(g,g)\,\big\}
\la{p1}
\end{equation} 
for any $\eps$ and any $\ell$, with the constant $k_\eps$ uniform in
$\ell,\L$. 
Once we have (\ref{p1}) we obtain Theorem \ref{teorema1}
by choosing $g:= E_s f$ and optimizing over $\ell$.
Indeed, with this choice we have $\nu(fE_s f)=\nu(f,g) = \nu(g,g)$ 
since $f$ (and therefore $E_sf$) has zero mean. Moreover 
$\cE_\nu(E_sf,E_sf)\leq s\, \nu(fE_s f)$, so that (\ref{p1}) implies
$$
\nu(fE_s f) \leq k_\eps\big\{s\,\ell^\eps + \ell^{-2}\big\}\,,
$$
and the claim follows.  
\acapo
In order to prove (\ref{p1}) we need the following technical lemma.
In what follows $f$ is as in Theorem \ref{teorema1}.
\begin{Le}
\la{62}
There exists a constant $k$ depending on $f$ such that 
\be
\Var_\nu\big(\, \nu(f\tc N_{\Si_{\ell,H}})\,\big) \leq 
\frac{k}{\ell^2} \qquad \forall \, \ell \leq \frac{L}{2}
\end{equation}
\end{Le}
\proof
Without loss of generality we can assume that $\ell$ is so large that
the support of $f$ is contained in $\Si_{\ell,H}$. For notation
convenience we set $N_{\ell} := N_{\Si_{\ell,H}}$ and 
$\mu_\ell^\l := \mu_{\Si_{\ell,H}}^\l$. Using the result on
the equivalence of ensembles, see Theorem \ref{equiv}, we can safely
replace $\nu(f\tc N_{\ell}=m)\,\big)$ with its
grand-canonical average $F(m):=\mu_{\ell}^{\l(m,\ell)}(f)$, where
$\l(m,\ell)$ is such that
$\mu_{\ell}^{\l(m,\ell)}(N_{\ell})=m$.  Moreover, thanks to
Proposition \ref{canvar}, we can bound the canonical variance
w.r.t.\ $\nu$ by the grand canonical one w.r.t.\ $\mu :=
\mu_\L^{\l(n)}$ with self explanatory notation. In conclusion
\be
\Var_\nu\big(\, \nu(f\tc N_{\ell})\,\big) \le
k\,\mu\big(F,F\big) + C \frac1{\ell^2} 
\end{equation}
for some $k=k(q)$ and $C=C(f,q)$. 
\acapo
Since the measure $\mu$ is a product measure over the sites
of $\L$, it is immediate to check (see also (\ref{poincare}))
\be
\mu\big(F,F\big) \le
\big\{\sum_{x\in \Si_{\ell,H}}\s^2_x\big\}\, 
\mu\big(2[F(N_\ell+1)-F(N_\ell)]^2 + 2[F(N_\ell)-F(N_\ell-1)]^2\big)
\la{A1}
\end{equation} 
where $\s^2_x := \mu(\a_x,\a_x)$.\acapo
We now bound the gradient $[F(m+1)-F(m)]$. Let $\l_s:= s\l(m+1,\ell)
+(1-s)\l(m,\ell)$ and let $F_s :=\mu_{\ell}^{\l_s}(f)$. Then, 
setting $a(q)=\log{1/q}$ we have
\begin{gather}
F(m+1)-F(m) = \int_0^1 \dd s\, \frac{\dd}{\dd s} F_s  
= a(q)\int_0^1\dd s\, \mu_{\ell}^{\l_s}(N_{\ell},f)
                                [\l(m+1,\ell)-\l(m,\ell)]
\la{der}
\end{gather}
In turn
\be
\l(m+1,\ell)-\l(m,\ell) = \int_m^{m+1} \dd t \,\frac{\dd}{\dd t}\l(t,\ell)
 = \int_m^{m+1}\dd t\, \big[a(q)\mu_{\ell}^{\l(t,\ell)}
      (N_{\ell},N_{\ell})\big]^{-1}  
\la{lambda}
\end{equation}
It is easy to check at this point, thanks to the results of \S 3.1, that
$|\l(m+1,\ell)-\l(m,\ell)|\leq k (m\wedge \ell^2)^{-1}$ for some
$k=k(q)$.  Since $|\mu_{\ell}^{\l_s}(N_{\ell},f)| \le
C_f$ we get that the r.h.s. of (\ref{der}) is bounded from above by
$C_fk (m\wedge \ell^2)^{-1}$. 
\acapo
In conclusion, the r.h.s. of (\ref{A1}) is bounded from above by
\be
K_f \ell^2 \big(\frac{n}{L^2} \wedge 1\big)
    \mu\Big(\big(N_{\ell}\wedge \ell^2\big)^{-2}\Big) 
\la{final}
\end{equation}
for some constant $K_f$ depending on $f$. Standard large deviations for
the product measure $\mu$ imply that the r.h.s. of (\ref{final}) is
bounded from above by $K'_f \ell^{-2}$.
\qed

\smallskip

We can now complete the proof of the theorem following step by step the
proof of Theorem 2.4 in \cite{CancMart}. We first establish (\ref{p1})
for $\eps = 2$ and then show how to improve it to all values of $\eps
>0$. \acapo The main ingredients are the lower bound on the spectral gap
given in Theorem \ref{teorema} together with the formula
$$
\nu(g,f) = \nu\big(\,\nu(g,f\tc \cF)\,\big) + \nu\big(\,g, \nu(f\tc \cF)\,\big)
$$
valid for any $\s$--algebra $\cF$. 
If we take $\cF$ as the $\s$--algebra generated by $N_\ell$, we get,
after one Schwartz inequality,
\begin{align}
\nu(g,f)^2 &\leq 2  \nu\big(\,\nu(g,f\tc N_\ell)^2\,\big) + 
               2  \nu\big(\,g, \nu(f\tc N_\ell)\,\big)^2
      \nonumber \\
    &\leq C_f \Big[\,\ell^2 \Dir_\nu(g,g) + \frac{1}{\ell^2}\Var_\nu(g)\,\Big] 
  \la{p11}
\end{align}
where we used Lemma \ref{62} and 
the Poincar\'e inequality $\Var_\nu(g\tc N_\ell) \leq k\,
\ell^2 \,\Dir_\nu(g,g\tc N_\ell)$, which follows from
Theorem \ref{teorema}.  
\acapo
Now we assume inductively that we have been able to prove (\ref{p11})
with $\ell^2$ replaced by $\ell^\eps$ and $C_f$ replaced by some
constant $C_{f,\eps}$ for all $\ell \le
\frac{L}{2}$. Then the term $\nu(g,f\tc N_\ell)^2$ in the r.h.s. of the
first line of (\ref{p11}) can be bounded from above by
\begin{align*}
\nu(g,f\tc N_\ell)^2 &\leq 
C_{f,\eps}\Big[\, \ell_1^\eps \Dir_\nu(g,g \tc N_\ell) +
               \frac{1}{\ell_1^2}\Var_\nu(g \tc N_\ell)\,\Big] \\
                     &\leq 
C'_{f,\eps}\Big[\, \ell_1^\eps  +
               \frac{\ell^2}{\ell_1^2}\,\Big]\Dir_\nu(g,g \tc N_\ell) 
\end{align*}
for any $\ell_1 \leq \frac{\ell}{2}$. If we optimize over $\ell_1$ for a
given $\ell$ we get   
\be
\nu(g,f\tc N_\ell)^2 \leq C''_{f,\eps} \ell^{\frac{2\eps}{2+\eps}}
                          \Dir_\nu(g,g \tc N_\ell)
\la{p12}
\end{equation}
In other words we have been able to replace the assumed $\ell^\eps$
factor in front of the Dirichlet form of $g$ by
$\ell^{\frac{2\eps}{2+\eps}}$. The price is an increase of the constant
$C_{f,\eps}$. Since the discrete map $x \rightarrow
\frac{2x}{2+x}$, $x_0=2$ has as unique fixed point the origin,
(\ref{p1}) follows.  \qed





%
%
%
%
%
\end{document}